\documentclass{aa}  

\usepackage{graphicx}
\usepackage{txfonts}
\usepackage{lscape}
\usepackage{hyperref}
\newcommand{\cobold}{{\sf CO$^5$BOLD\ }}
\newcommand{\asset}{{\sf ASS$\epsilon$T}}

\begin{document}

\title{Solar center-to-limb variation in Rossiter-McLaughlin and exoplanet
  transmission spectroscopy}

\titlerunning{Solar CLV in transmission and RM curves}
   
   \author{A. Reiners\inst{1}
          \and
          F. Yan\inst{2}
          \and
          M. Ellwarth\inst{1}
          \and
          H.-G. Ludwig\inst{3}
          \and
          L. Nortmann\inst{1}
        }
        \institute{
          Institut f\"ur Astrophysik und Geophysik, Georg-August-Universit\"at, D-37077 G\"ottingen, Germany
          \and
          Department of Astronomy, University of Science and Technology of
          China, Hefei 230026, China
          \and
          Landessternwarte, Zentrum f\"ur Astronomie der Universit\"at Heidelberg,
          D-69117 Heidelberg, Germany
        }
   \date{\today}

   \abstract{Line profiles from spatially unresolved stellar observations
     consist of a superposition of local line profiles that result from
     observing the stellar atmosphere under specific viewing angles. Line
     profile variability caused by stellar magnetic activity or planetary
     transit selectively varies the weight and/or shape of profiles at
     individual surface positions. The effect is usually modeled with
     radiative transfer calculations because observations of spatially
     resolved stellar surfaces are not available. For the Sun, we recently
     obtained a broadband spectroscopic atlas of the solar center-to-limb
     variation (CLV). We use the atlas to study systematic differences between
     largely used radiative transfer calculations and solar observations. We
     concentrate on four strong lines useful for exoplanet transmission
     analysis, and we investigate the impact of CLV on transmission and
     Rossiter-McLaughlin (RM) curves. Solar models used to calculate synthetic
     spectra tend to underestimate line core depths but overestimate the
     effect of CLV. Our study shows that CLV can lead to significant
     systematic offsets in transmission curves and particularly in RM curves;
     transmission curves centered on individual lines are overestimated by up
     to a factor of two by the models, and simulations of RM curves yield
     amplitudes that are off by up to 5--10\,m\,s$^{-1}$ depending on the
     line. For the interpretation of transit observations, it is crucial for
     model spectra that accurately reproduce the solar CLV  to become available
     which, for now, is the only calibration point available.}

   
   \keywords{keywords}

   \maketitle
%

\section{Introduction}

The solar spectrum varies across the visible surface because of limb darkening
caused by temperature stratification of the solar atmosphere, and because the
opacity distribution and non-local thermodynamic equilibrium (non-LTE) effects
differ between spectral lines \citep[see, e.g.,][]{2015A&A...573A..74S}. The
projected convective velocity distribution depends on the formation height and
angle of projection. Limb darkening and its wavelength variability provide
observable constraints for the structure of the Sun's atmosphere
\citep{Asplund2009} which, until today, have only partially been
exploited. One reason for this is the lack of observations at different values
of center-to-limb distance, parameterized by $\mu = \cos{\theta}$ with the
heliocentric angle~$\theta$. \citet{Ellwarth2022} recently presented a
spectral library of solar center-to-limb variation (CLV).

The Sun serves as a benchmark for other stars. Doppler radial velocity (RV)
measurements and transit spectroscopy critically depend on differential
comparison between stellar reference or template spectra, and spectroscopic
data taken at different times. Most stars other than the Sun cannot be
spatially resolved, and assumptions about their limb darkening and variability
are motivated by the solar example. An important task for improving RV
measurement precision is understanding how RV variability is driven by
convection and its interplay with magnetic fields \citep{2010A&A...512A..38L,
  2010A&A...520A..53L, 2010A&A...512A..39M, 2010A&A...519A..66M,
  2014ApJ...796..132D, 2015ApJ...798...63M, 2016MNRAS.457.3637H,
  2019APJ...874..107M}. The motion of hot, rising plasma on the surface of a
star causes a Doppler shift of the observed spectrum
\citep{1978soph...58..243b, 1981A&A....96..345d, 1982natur.297..208l}. Because
the motion of the plasma is nonisotropic, the amplitude of the Doppler shift
varies across the stellar disk, which means it depends on the stellar limb
position, and can be visualized spectroscopically \citep{1984soph...93..219b,
  1985A&A...150..256c}. For a Sun-like star, the disk-integrated effect is on
the order of a hundred m\,s$^{-1}$ such that the observed RV of a star is
systematically blueshifted \citep[see, e.g.,][]{2009LRSP....6....2N}.

Atmospheres of extrasolar planets can be studied during planetary transits
when the light of the star penetrates the planetary atmosphere. For this, the
background stellar spectrum must be known to very high precision, and the
contribution from the part of the surface occulted during transit needs to be
carefully considered. Knowledge about the nonuniformity of the stellar flux
across the stellar disk is crucial for this technique
\citep{2015A&A...582A..51C, 2017A&A...603A..73Y}. Differential limb darkening
affects the influence of the transiting planet on the spectrum and is a major
source of uncertainty for transit spectroscopy \citep{2013A&A...549a...9c,
  2015mnras.450.1879e, 2015mnras.453.3821p, 2019A&A...622a..33m}. Furthermore,
the stellar lines have very different shapes across the stellar surface
\citep[the effect of CLV;][]{1972soph...24...18a, 1981A&A....96..345d,
  1984soph...93..219b, 2015A&A...574a..94y}. Recently, the presence of
\ion{Na}{i} and other absorbing species in the atmosphere of HD~209458b was
questioned by transit observations with different instruments
\citep{2020A&A...635A.206C, 2021A&A...647A..26C}. Features around the stellar
\ion{Na}{i} lines observed during transit could be interpreted as signatures
of the deformation of the stellar line profiles, due to the
Rossiter-McLaughlin (RM) effect and CLV. The signature is very similar to the
one from exoplanet atmosphere absorption.

\citet{2019A&A...631a.100c} demonstrate that considering limb darkening and
CLV significantly improves planet detectability from transit
spectroscopy. This, however, must go together with improving our understanding
of stellar magnetoconvection. Unfortunately, no suitable observations of
spatially resolved stellar surfaces are available against which the model line
profiles across the stellar disk could be calibrated.  In this context,
\citet{2017A&A...605A..90D} point out the importance of testing 3D and
time-dependent hydrodynamic models, as the one used in this work, against
observations, and \citet{2021A&A...649a..17d} discuss the prospects for
obtaining spatially resolved stellar spectroscopy from transit
observations. \citet{2017A&A...605A..91D} attempted to extract local line
profiles from transit observations of HD~209458, and solar eclipse
observations by \citet{2016A&A...595A..26R} demonstrate the same conceptual
difference between a model and observations. An improvement in quality is
expected for observations in stars other than the Sun when observations with
30\,m-class telescopes become available. Until then, the Sun is the only star
for which spatially resolved spectroscopy at the required spectral fidelity is
feasible.

We recently obtained a new grid of the spectrally resolved solar CLV
\citep{Ellwarth2022}. In this paper, we select four spectral lines and
demonstrate data quality and line profile variations at different limb
positions. We compare the lines to models of line profiles from simulations
that are frequently used for the modeling of RV variability and transit
observations, and we estimate the impact on transmission curves and the RM
effect.

\section{Data}

We compare observations of the solar CLV with three sets of model spectra. All
datasets are accessible
online.\footnote{\url{http://www.astro.physik.uni-goettingen.de/research/solar-lib/}}

\subsection{Observations}

Our observations of the Sun at different limb positions, $\mu=\cos{\theta}$,
are described in detail in \citet{Ellwarth2022}. We obtained spectra at 14
different limb positions of which we use the following 12 for this paper:
$\mu = 0.2, 0.3, 0.4, 0.5, 0.6, 0.7, 0.8, 0.9, .95, 0.98, 0.99,$ and $ 1.0$.
With respect to the available 14 positions, we chose to neglect two at
$\mu = 0.35$ and 0.97 to achieve a more regular sampling.  The Sun was
observed with the 50\,cm Vacuum Vertical Telescope at the Institute for
Astrophysics and Geophysics, G\"ottingen \citep{2020SPIE11447E..A9S}. Spectra
were obtained with our Fourier Transform Spectrometer
\citep[FTS;][]{2020SPIE11447E..3QS} in the range $\lambda$ = 4200 -- 8000\,\AA\
at a resolving power $R = 700\,000$ at $\lambda = 6000\,\AA$
($\Delta\nu = 0.0024$\,cm$^{-1}$). Each observation covers the entire
wavelength range providing superior wavelength (frequency) calibration
\citep[see][]{2016A&A...587A..65R}. We estimate that the linearity of the
wavelength calibration is better than 0.5\,m\,s$^{-1}$ at all wavelengths
\citep{Huke:19} and that the uncertainty of the absolute zero point offset of
each spectrum is below 30\,m\,s$^{-1}$ \citep{Ellwarth2022}.

The CLV atlas was not corrected for gravitational redshift. We applied a Doppler
shift of $\Delta\varv=-633.1$\,m\,s$^{-1}$ \citep{2020A&A...643A.146G} for
the comparison with model spectra.

\subsection{Synthetic spectra}

In the following, we compare our observations to three sets of synthetic
spectra to demonstrate typical model predictions about CLV. Our goal is to
identify the typical systematic differences between models and the solar
spectrum. We concentrate on ``standard'' models that are commonly used in
stellar line profile analysis and exoplanet transmission spectroscopy. We do
not attempt to provide the best match possible but calculate the model spectra
with parameters according to a ``typical'' main sequence Sun-like star.

Two of the fundamental choices in computations of spectral line formation are
the treatment of local thermodynamic equilibrium (LTE) and of convection. We
computed 1D models for the assumption of LTE and with corrections for NLTE,
and we computed spectra with a 3D convection model under the assumption of
LTE. We leave computations of 3D convection together with NLTE and
optimization of any parameters for in-depth studies of atmospheric models. An
overview of the model parameters is provided in Table\,\ref{tab:models}. As
for the expected differences between the 1D and 3D approach, and the
difference between LTE and NLTE spectral line formation, we refer to
\citet{2009LRSP....6....2N} and \citet{Asplund2009} and the SME references
(see below).

\begin{table}
  \centering
  \caption{Parameters of the synthetic calculations.}
  \label{tab:models}
  \begin{tabular}{lccc}
    \hline\hline\noalign{\smallskip}
    & SME LTE & SME NLTE & \cobold \\
    \hline\noalign{\smallskip}
    LTE/NLTE & LTE & NLTE & LTE \\
    1D/3D & \multicolumn{2}{c}{1D} & 3D \\
    $T_{\rm eff}$ (K) & \multicolumn{2}{c}{5800} & 5774 \\
    $\log{g}$ & \multicolumn{2}{c}{4.45} & 4.44 \\
    $\lambda$ range (\AA) & \multicolumn{2}{c}{4700 -- 7900} & 1200 -- 10000 \\
    $\mu$ values & \multicolumn{2}{c}{[0.2, 0.3, 0.4, \dots, 0.9,} & [0.09, 0.41, \\
    & \multicolumn{2}{c}{0.95, 0.98, 0.99, 1.0]} &  0.79, 1.0] \\
    sampling ($\Delta \varv$) & \multicolumn{2}{c}{130\,m\,s$^{-1}$} & 330\,m\,s$^{-1}$\\   
    \hline\noalign{\smallskip}
  \end{tabular}
\end{table}

\subsubsection{SME}

One-dimensional models in LTE and NLTE were computed with the software package
Spectroscopy Made Easy \citep[SME;][]{1996A&AS..118..595V,
  2017A&A...597A..16P}\footnote{\url{https://www.stsci.edu/~valenti/sme.html}}
with atomic and molecular line data from VALD3 \citep{1995A&AS..112..525P,
  1999A&AS..138..119K} and MARCS atmospheric models
\citep{2008A&A...486..951G}. For effective temperature and surface gravity, we
chose $T_{\textrm{eff}} = 5800$\,K and $\log{g} = 4.45$, and solar abundances
are adopted from \citet[][see
Appendix\,\ref{app:abundances}]{2007SSRv..130..105G}. Microturbulence was set
to $\varv_{\textrm{mic}} = 1$\,km\,s$^{-1}$, and macroturbulence was set to
zero. Departure coefficients for NLTE calculations were taken from
\citet{2016MNRAS.463.1518A} for Fe and from \citet{2020A&A...642A..62A} for
Ca, K, Mg, Na, and O. We computed a grid of models for the same $\mu$ values
as our observations.

\subsubsection{\cobold}

A synthetic LTE spectrum was calculated with the \asset\ code
\citep{2008ApJ...680..764K,2009AIPC.1171...73K} for a 3D and
time-dependent solar model atmosphere calculated with the \cobold\ code
\citep{2012JCoPh.231..919F}. The code explicitly includes and resolves
convective motions in the atmosphere. Hydrodynamical simulations of the stellar
convection code with \cobold\ solve the equations for the compressible
hydrodynamics coupled with non-local transport of radiation with detailed
opacities. The line spectral line list from the TLUSTY/SYNSPEC code was used
\citep{2011ascl.soft09021H}. The particular \cobold\ model was previously
applied in projects aiming to determine several elemental
abundances in the solar photosphere
\citep{2011SoPh..268..255C,2015A&A...583A..57S}, and it is a member of the {\sf
  CIFIST} grid of 3D model atmospheres
\citep{2009MmSAI..80..711L,2013A&A...557A...7T}. In brief, the computational
domain has a size of $5.6\times 5.6\times 2.25\,\mathrm{Mm}^3$ covered by
$140 \times 140 \times 150$ mesh points. The temporal evolution over 2.1\,h
was represented by 20 statistically independent instances in time. To reduce
computational cost, the model was horizontally subsampled by a factor of
three. The final spectra for the various limb positions are averages of the
emergent intensity over time and horizontal
position. Appendix\,\ref{app:abundances} lists the abundances that were
assumed in the spectral synthesis with \asset. We emphasize that no effort was
made to fit particular lines or spectral regions.

Synthetic \cobold\ spectra are available for limb positions
$\mu = 0.09, 0.41, 0.79,  1.0$. For comparison to observations, we interpolated
the \cobold\ spectra to the observed $\mu$ grid. We applied 2D
cubic convolution interpolation ($\lambda$, $\mu$) based on the triangulation
algorithm from \citet[][]{LeeSchachter80}.

\section{Center-to-limb variation}
\label{sect:CLV}

\begin{figure*}
  {\sffamily
  \centering
  \mbox{\parbox{.95\textwidth}{\tiny \hspace{28mm} \ion{Na}{i} \hspace{50mm} \ion{Na}{i}\hspace{50mm} \ion{Na}{i}}}\\[-5mm]
  \mbox{
    \parbox{.95\textwidth}{
      \resizebox{.3\textwidth}{!}{\includegraphics[viewport=50 25 690 690]{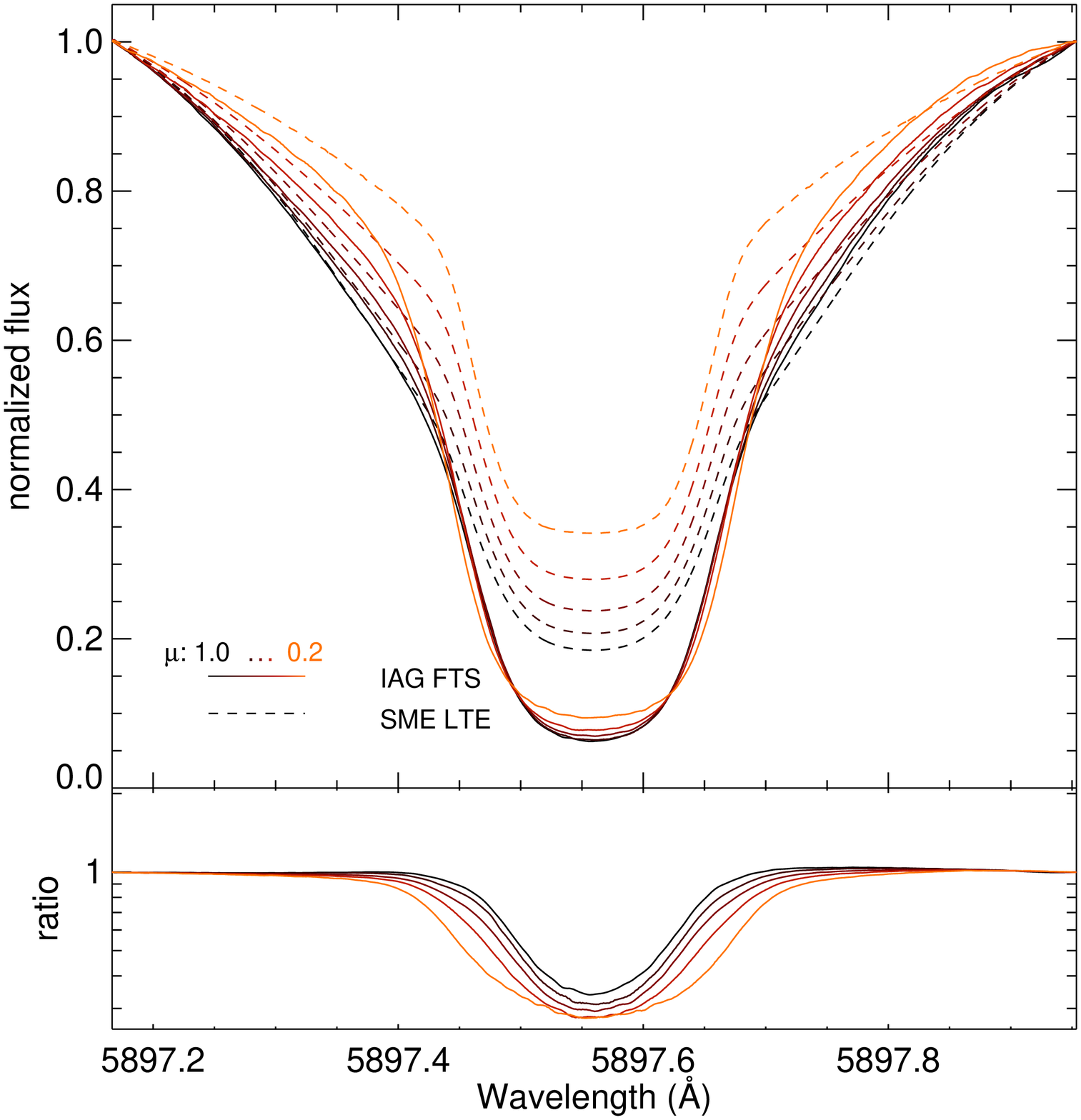}}
      \resizebox{.3\textwidth}{!}{\includegraphics[viewport=50 25 690 690]{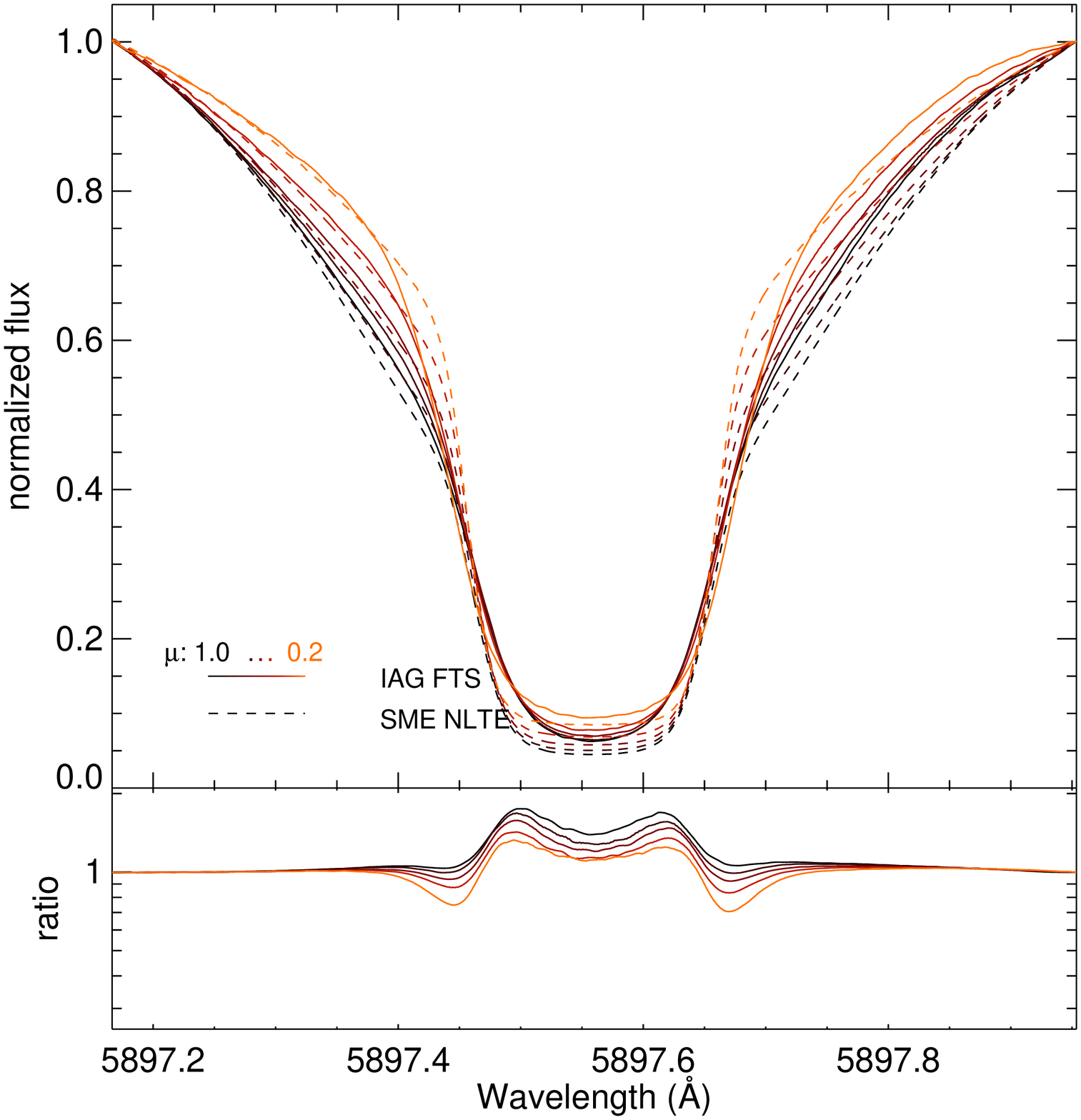}}
      \resizebox{.3\textwidth}{!}{\includegraphics[viewport=50 25 690 690]{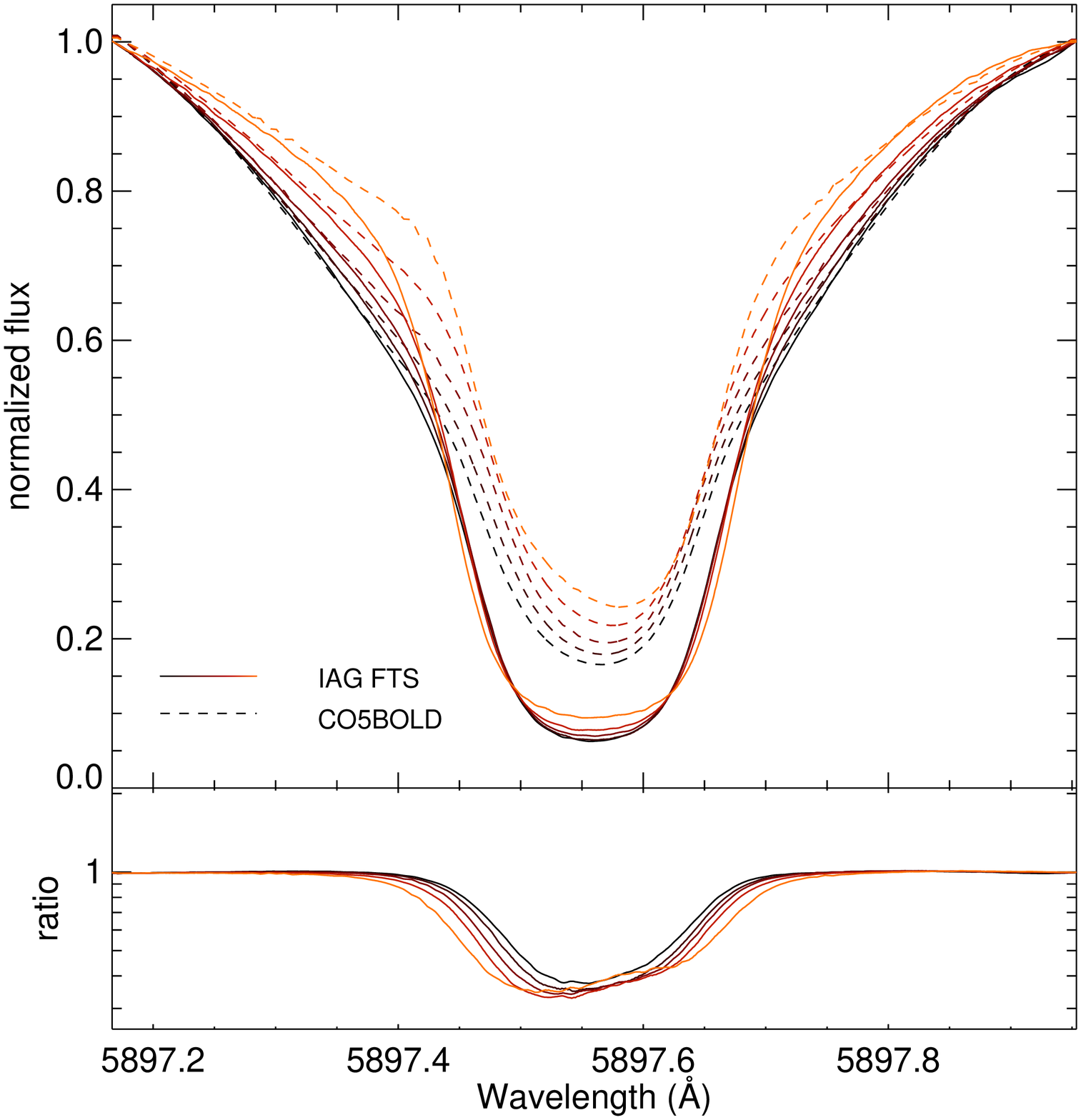}}
    }}
  \mbox{\parbox{.95\textwidth}{\tiny \vspace{3mm} \hspace{28mm} \ion{Mg}{i} \hspace{50mm} \ion{Mg}{i}\hspace{50mm} \ion{Mg}{i}}}\\[-5mm]
  \mbox{
    \parbox{.95\textwidth}{
      \resizebox{.3\textwidth}{!}{\includegraphics[viewport=50 25 690 690]{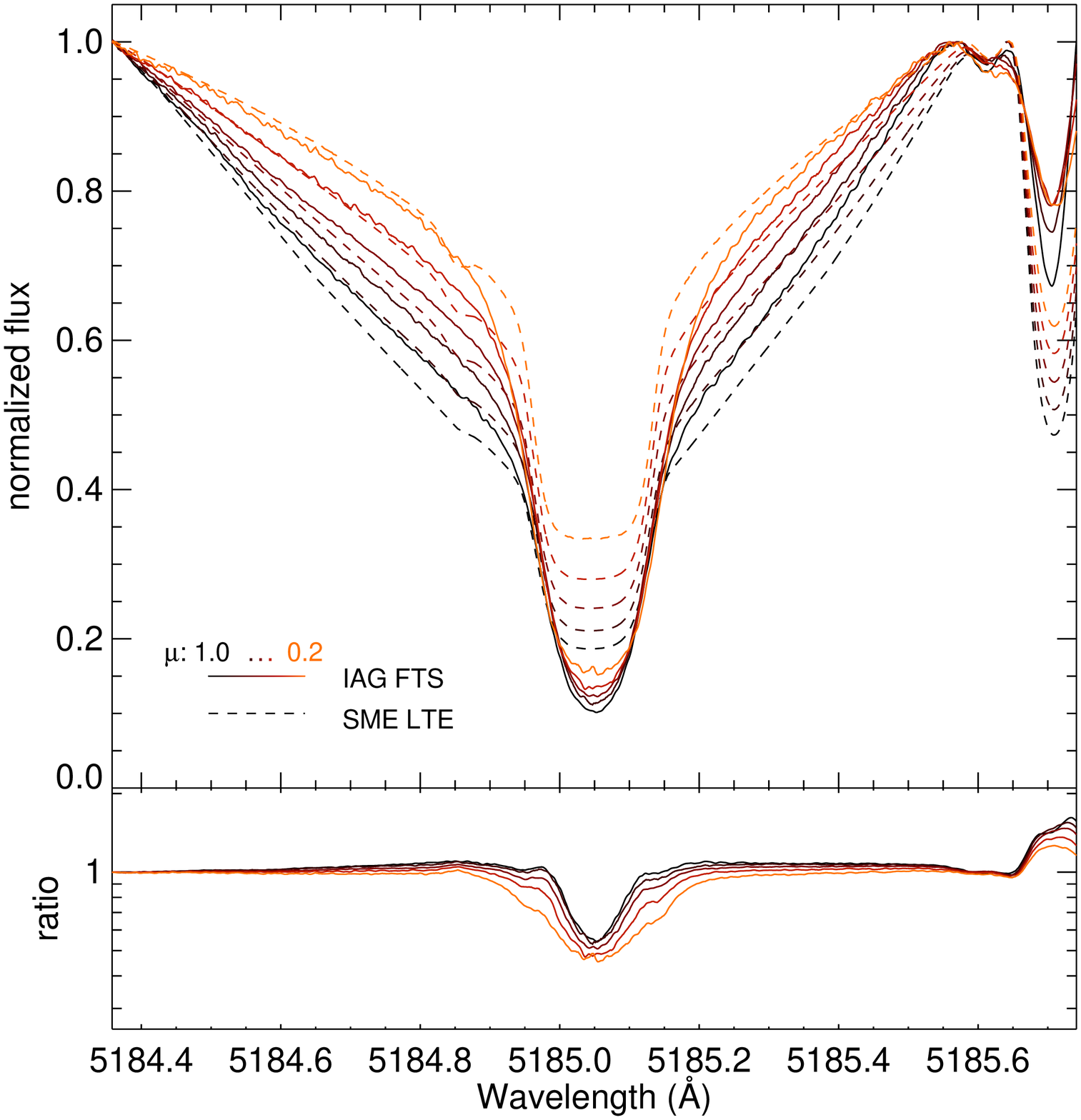}}
      \resizebox{.3\textwidth}{!}{\includegraphics[viewport=50 25 690 690]{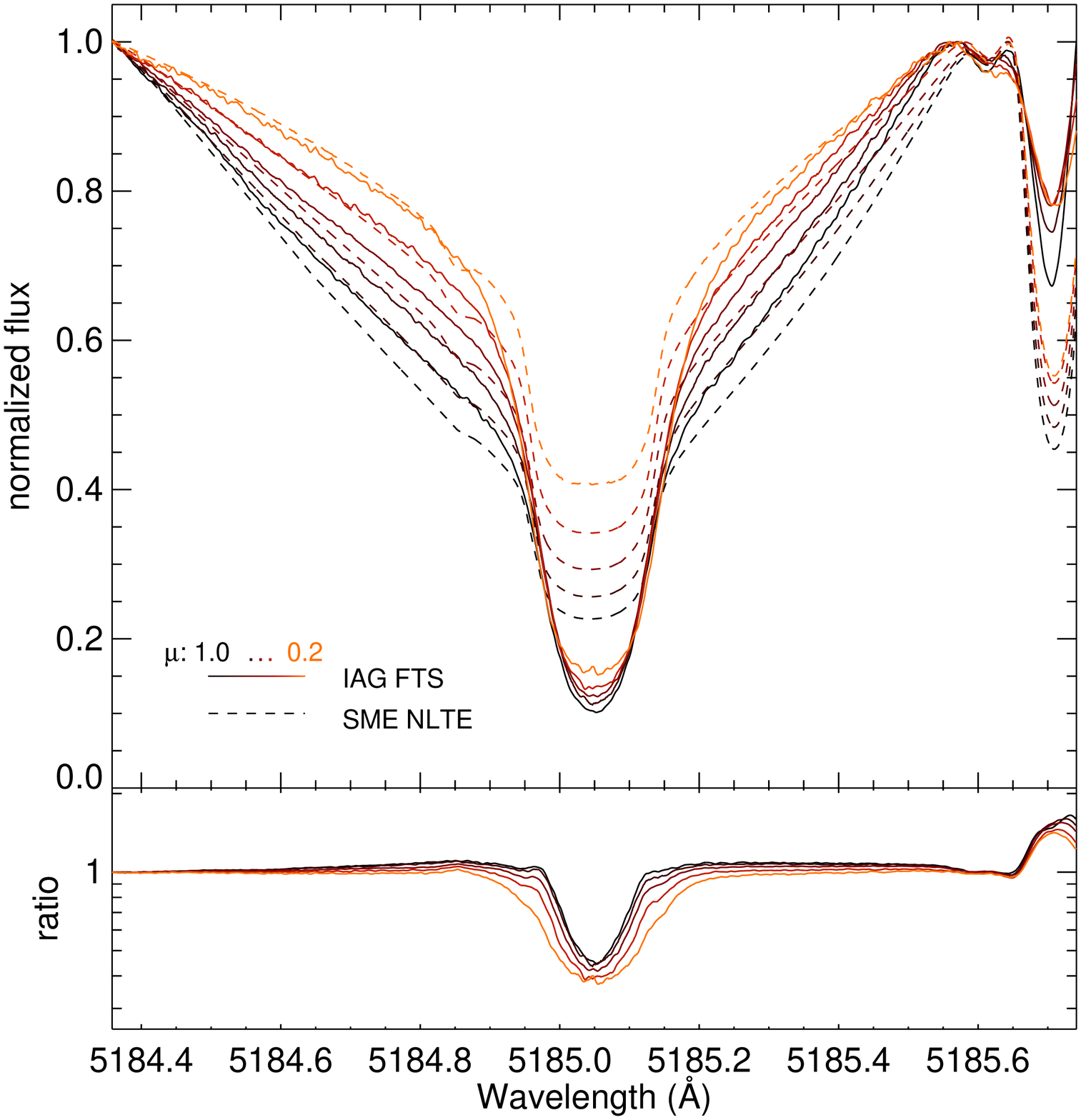}}
      \resizebox{.3\textwidth}{!}{\includegraphics[viewport=50 25 690 690]{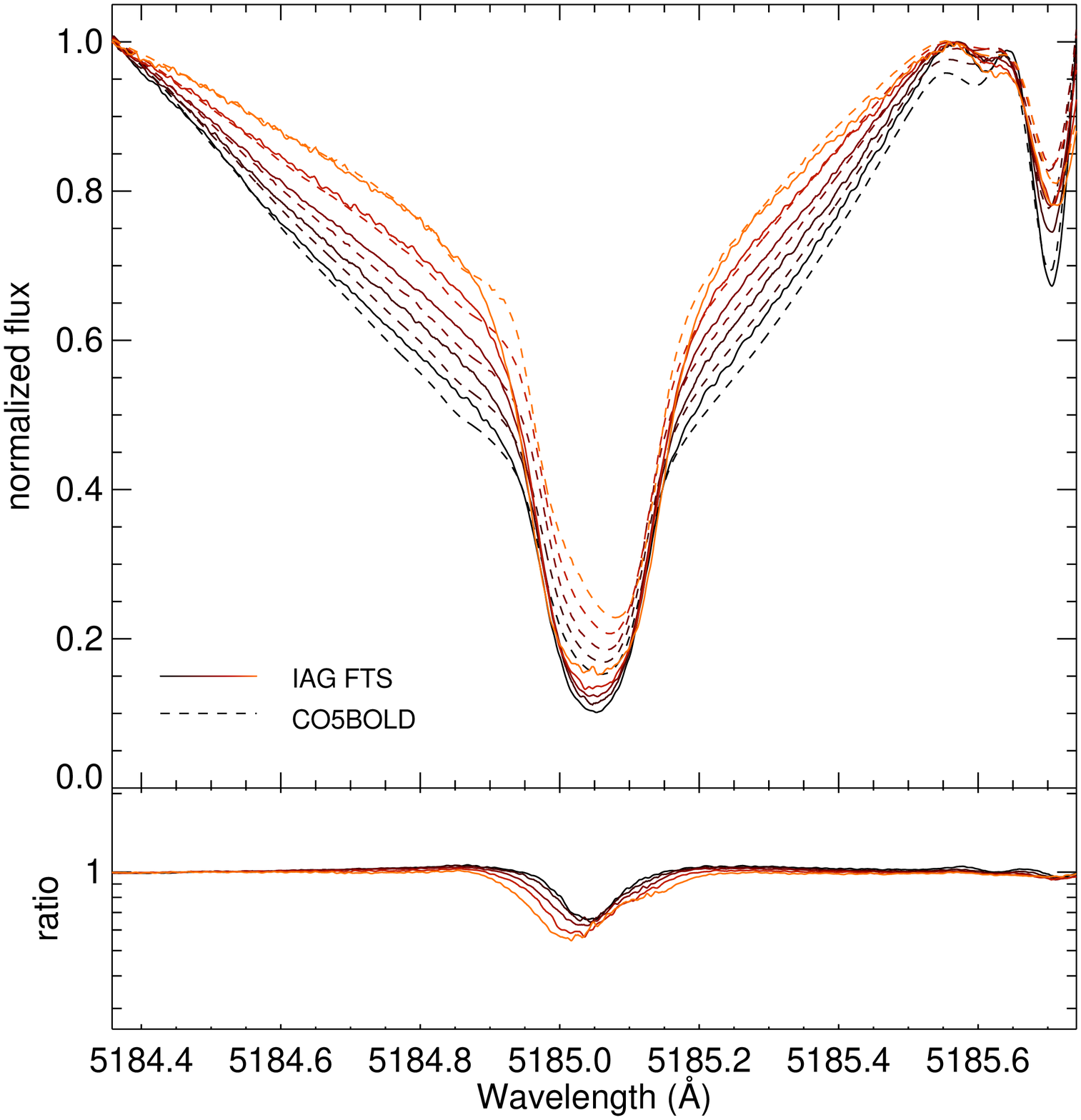}}
    }}
  \mbox{\parbox{.95\textwidth}{\tiny \vspace{3mm} \hspace{28mm} \ion{Ca}{i} \hspace{50mm} \ion{Ca}{i}\hspace{50mm} \ion{Ca}{i}}}\\[-5mm]
  \mbox{
    \parbox{.95\textwidth}{
      \resizebox{.3\textwidth}{!}{\includegraphics[viewport=50 25 690 690]{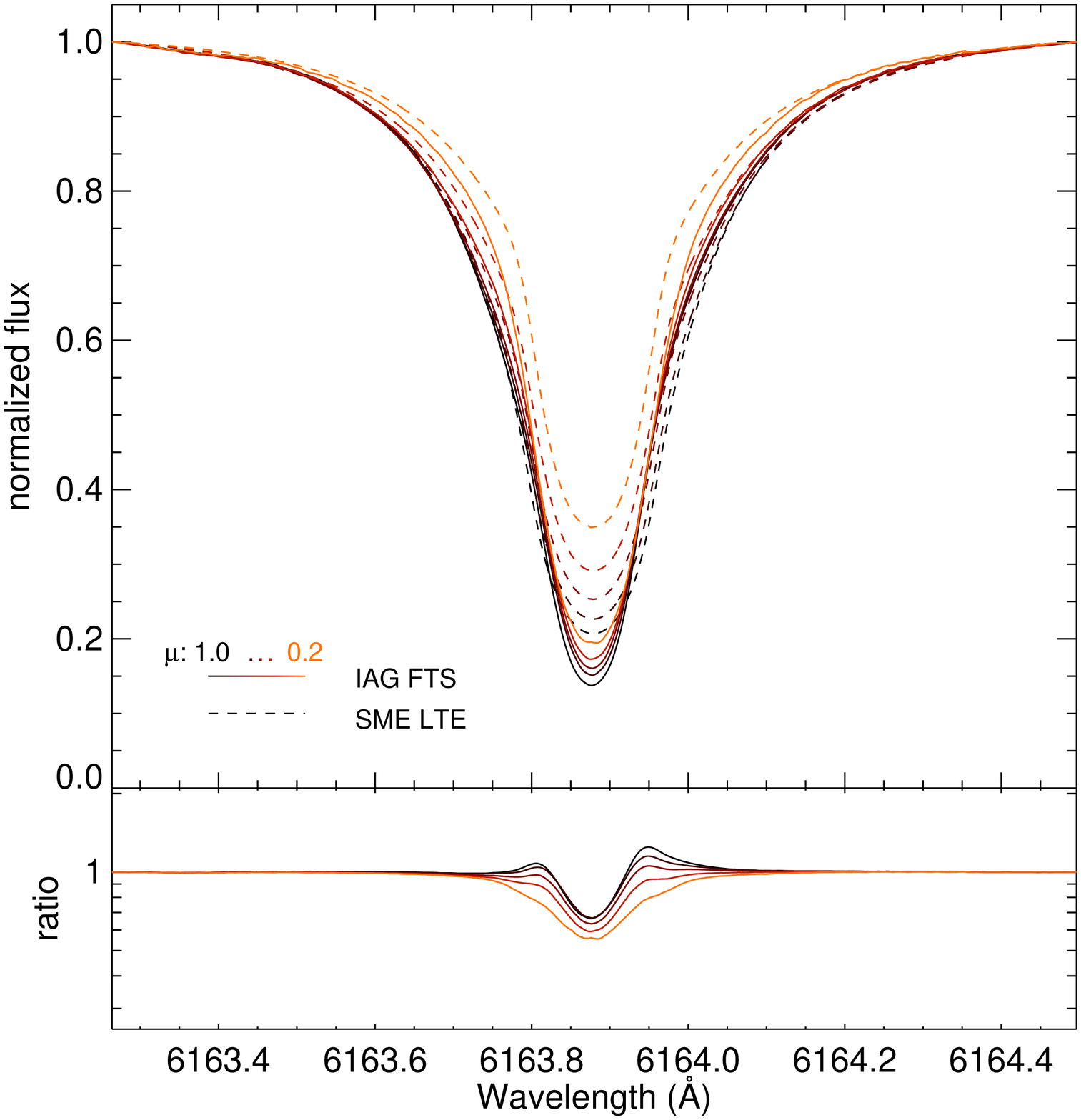}}
      \resizebox{.3\textwidth}{!}{\includegraphics[viewport=50 25 690 690]{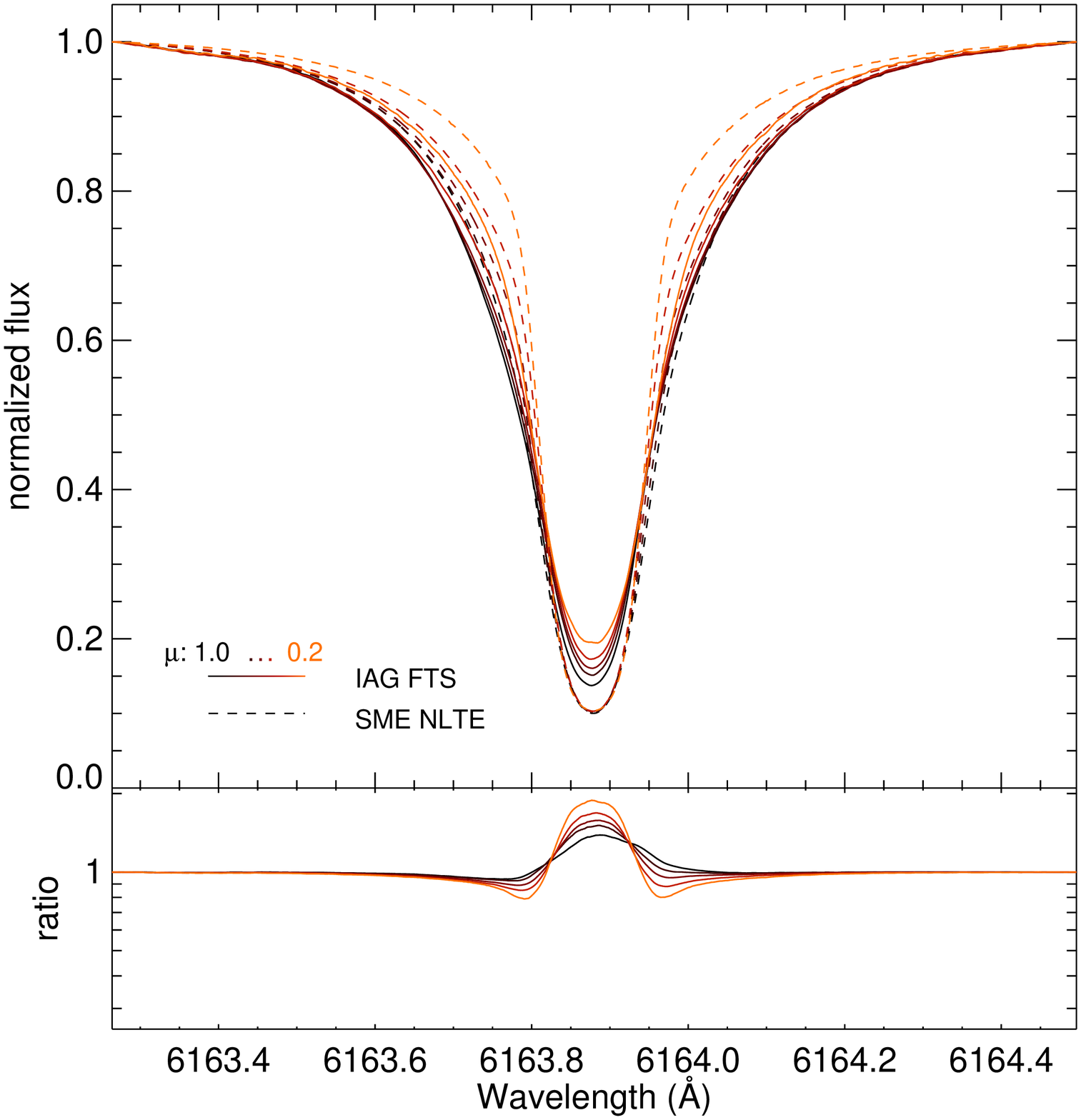}}
      \resizebox{.3\textwidth}{!}{\includegraphics[viewport=50 25 690 690]{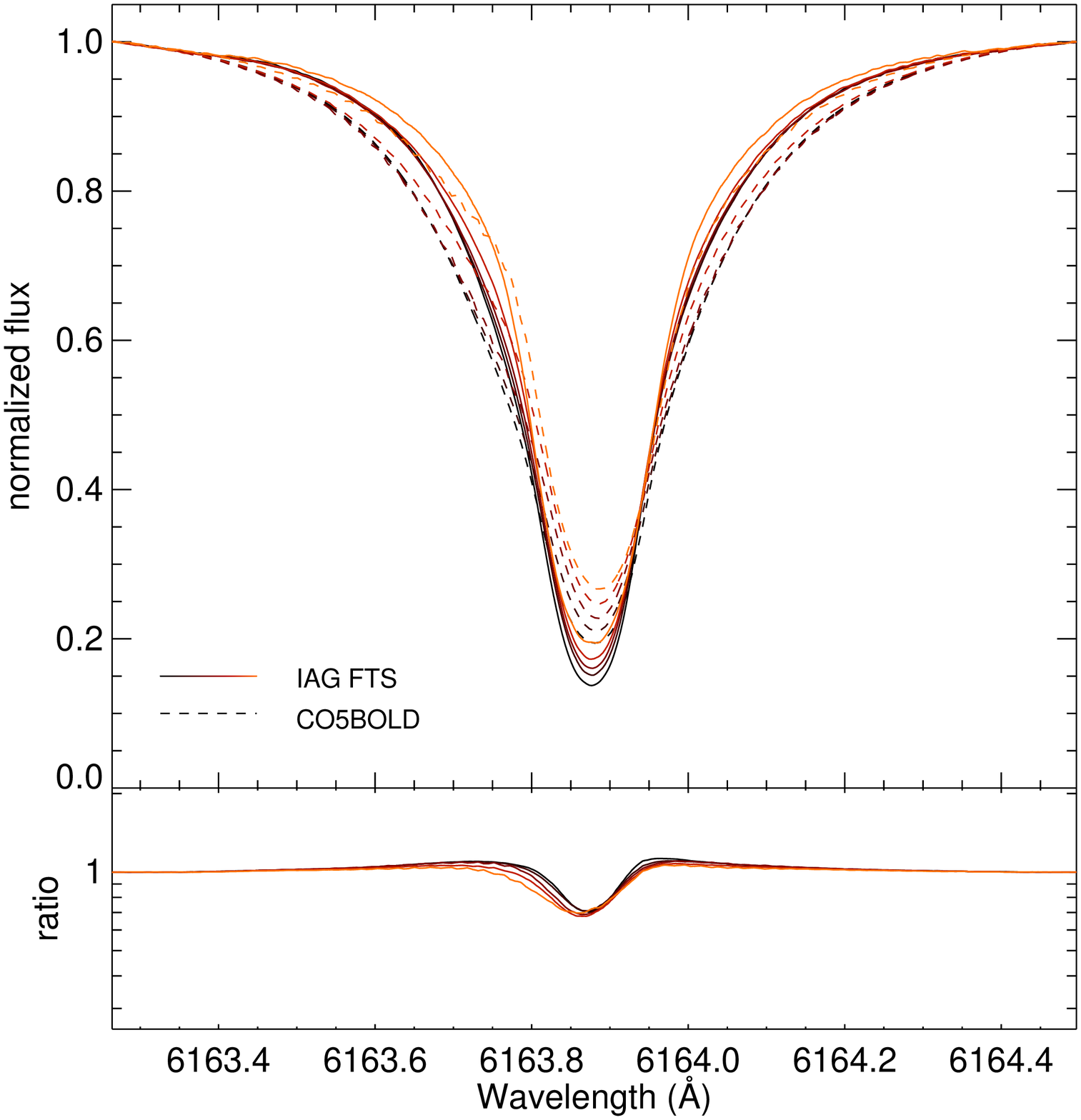}}
    }}
  \mbox{\parbox{.95\textwidth}{\tiny \vspace{3mm} \hspace{28mm} \ion{Fe}{i} \hspace{50mm} \ion{Fe}{i}\hspace{50mm} \ion{Fe}{i}}}\\[-5mm]
  \mbox{
    \parbox{.95\textwidth}{
      \resizebox{.3\textwidth}{!}{\includegraphics[viewport=50 25 690 690]{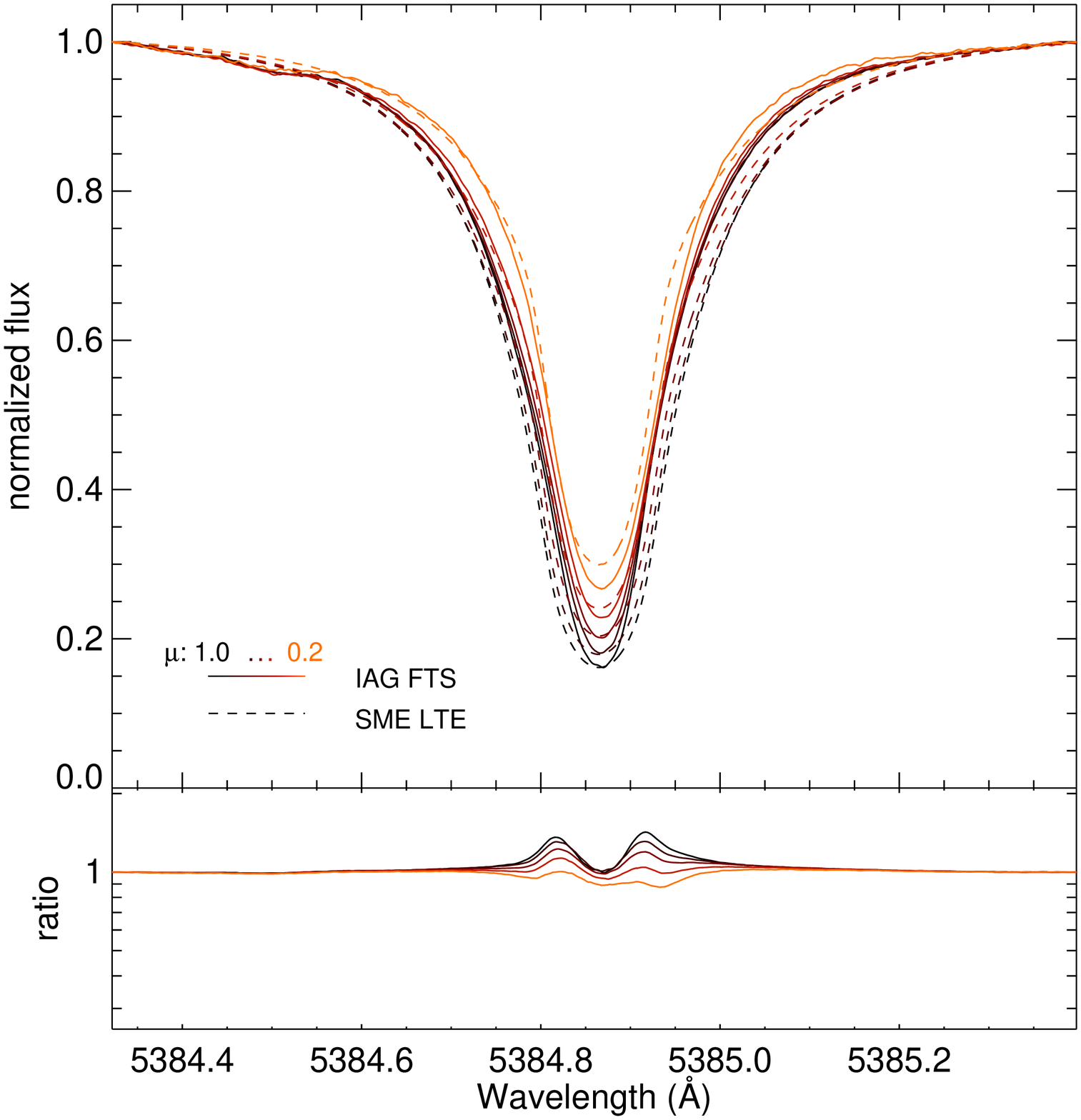}}
      \resizebox{.3\textwidth}{!}{\includegraphics[viewport=50 25 690 690]{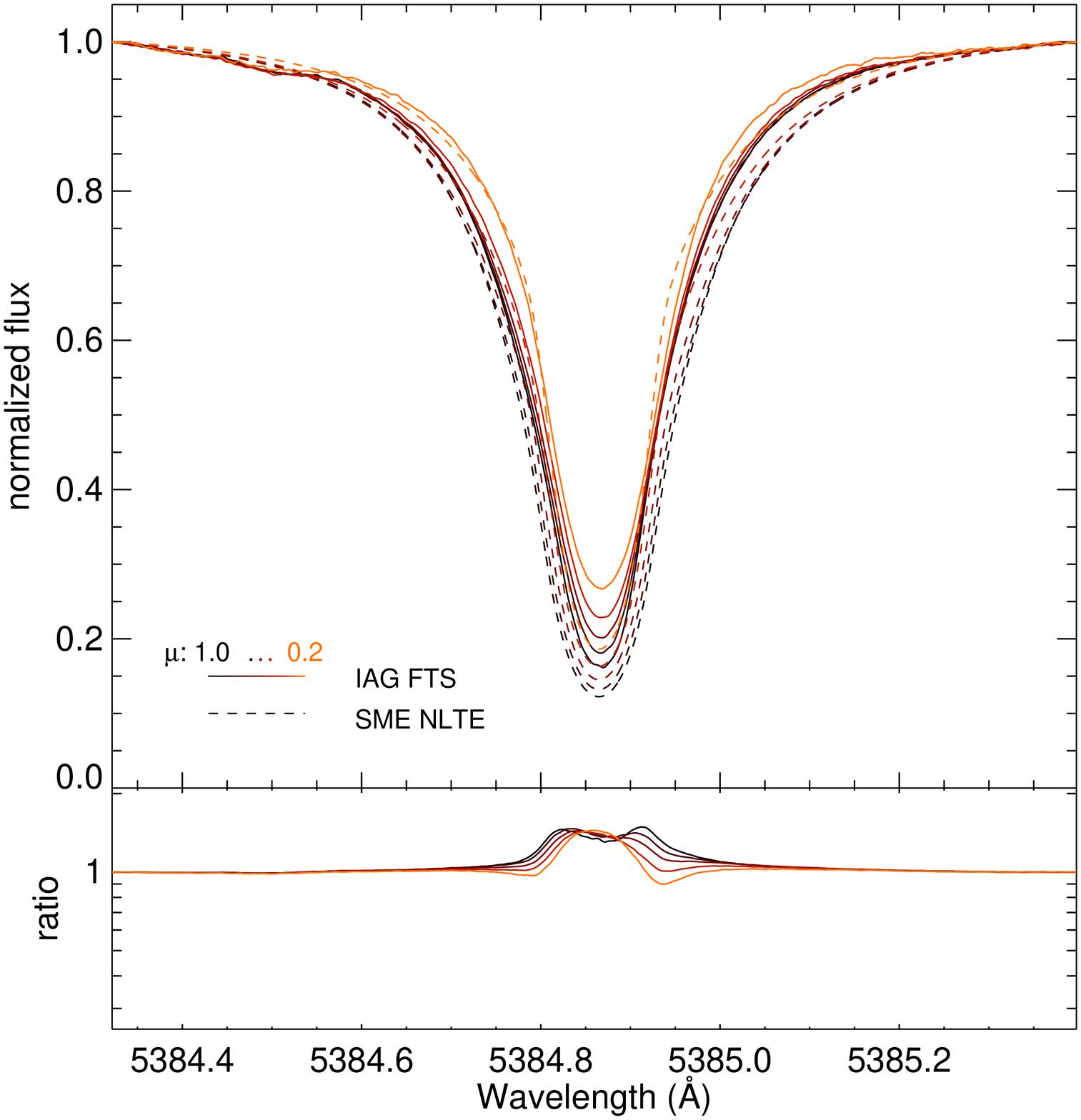}}
      \resizebox{.3\textwidth}{!}{\includegraphics[viewport=50 25 690 690]{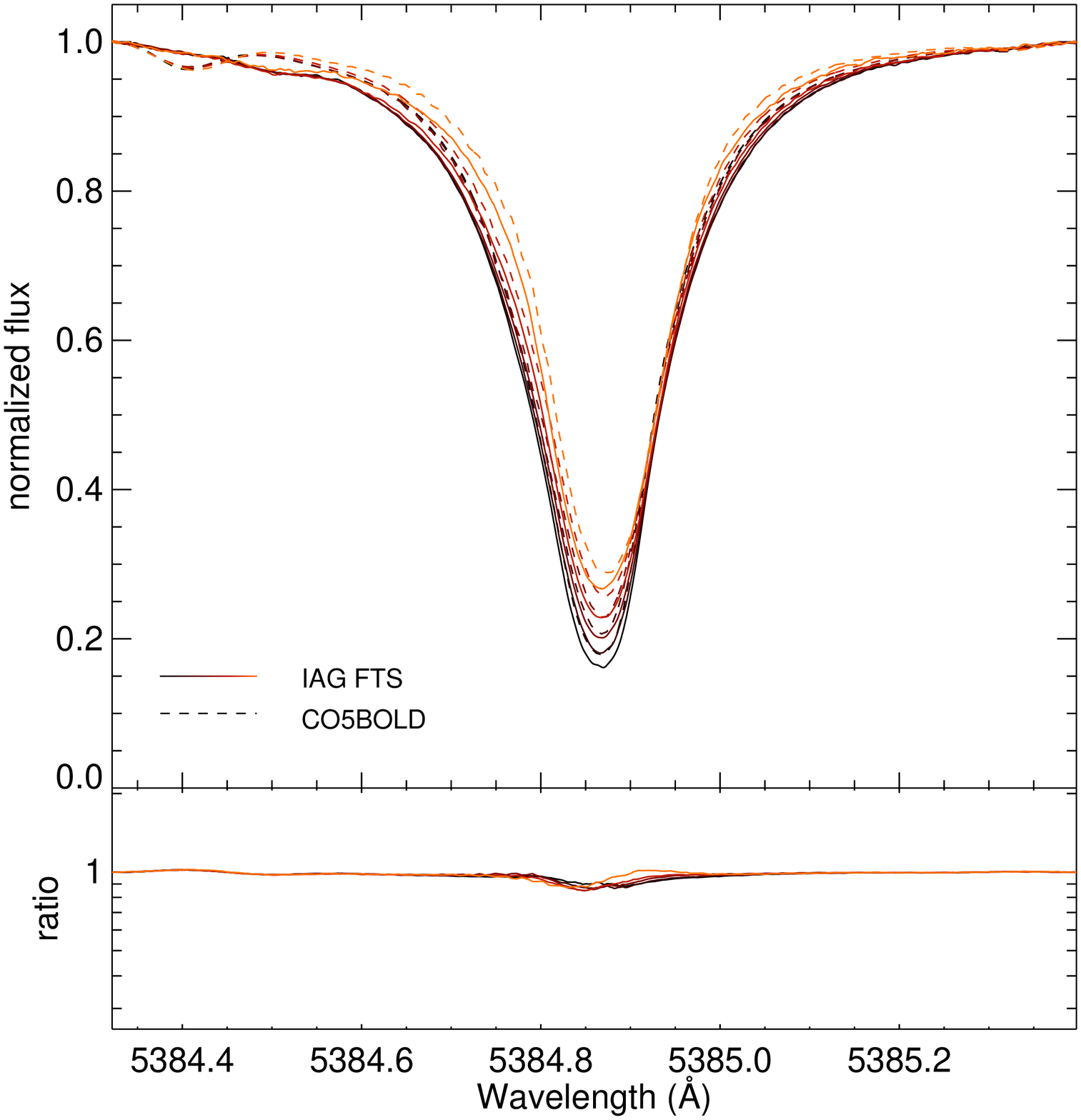}}
    }}
  \caption{\label{fig:CLV}Line profile comparison between observed and model
    spectra for four example lines \ion{Na}{I} 5897, \ion{Mg}{I} 5185,
    \ion{Ca}{I} 6163, \ion{Fe}{I} 5384 (top to bottom; panels span velocity ranges $\pm20, \pm40, \pm30$, and
    $\pm30$\,km\,s$^{-1}$, respectively), and for model spectra
    for SME~LTE, SME~NLTE, and \cobold calculations (left to right).  For each
    panel, the upper plot shows observed IAG FTS spectra as solid lines and
    model spectra as dashed lines. Lower plots show data/model ratios on a
    logarithmic scale. Five limb positions, $\mu = 0.2, 0.4, 0.6, 0.8, 1.0$, are
    shown for each set of spectra. Darker color is used for spectra taken
    closer to disk center. }
}
\end{figure*}

\begin{figure*}
  {\sffamily
  \centering
  \mbox{\parbox{.95\textwidth}{\tiny \hspace{31mm} \ion{Na}{i} \hspace{40mm} \ion{Na}{i} \hspace{40mm} \ion{Na}{i}\hspace{40mm} \ion{Na}{i}}}\\[-5mm]
  \mbox{
    \parbox{\textwidth}{
      \resizebox{.24\textwidth}{!}{\includegraphics[viewport=35 10 622 442]{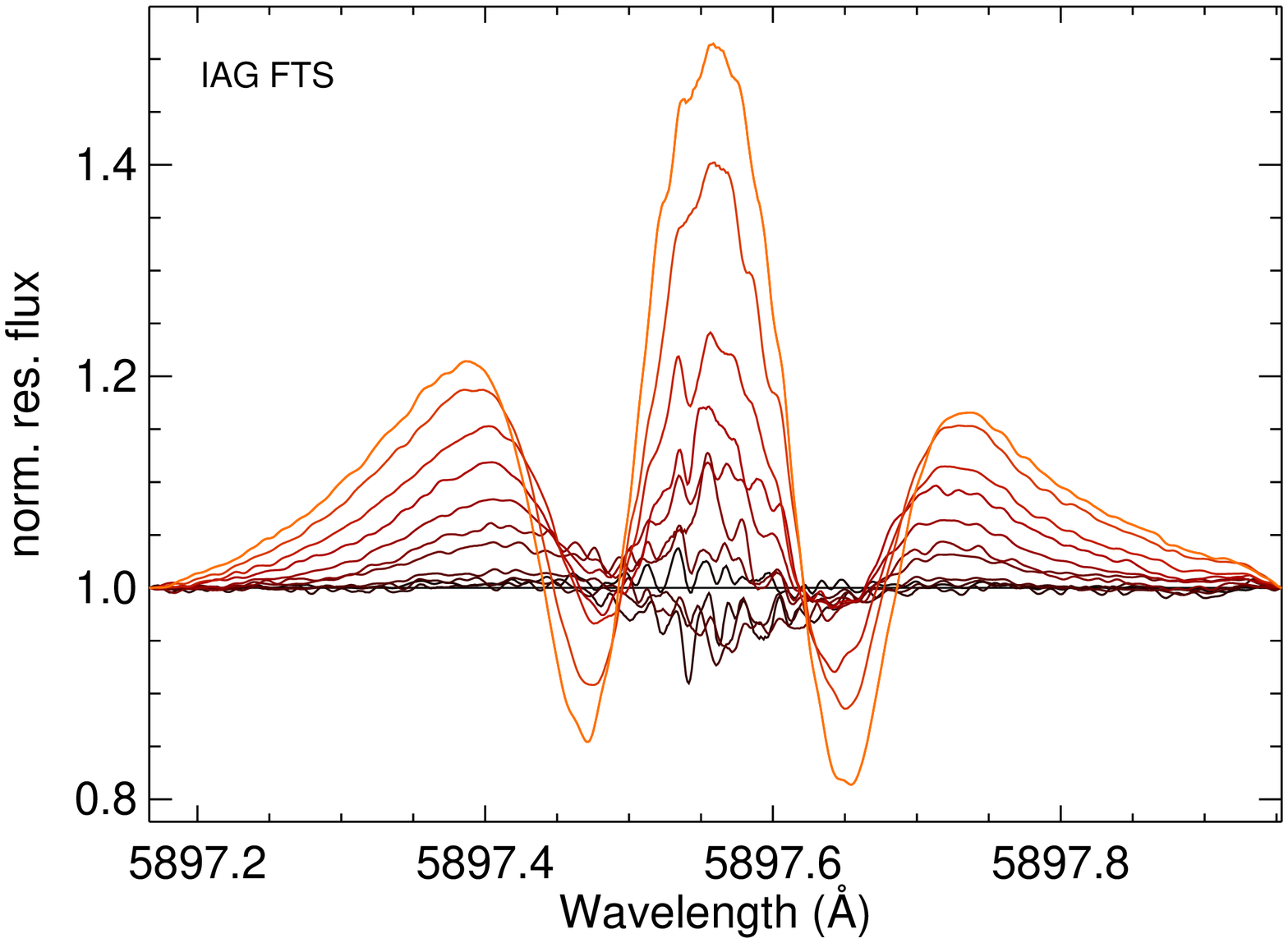}}
      \resizebox{.24\textwidth}{!}{\includegraphics[viewport=35 10 622 442]{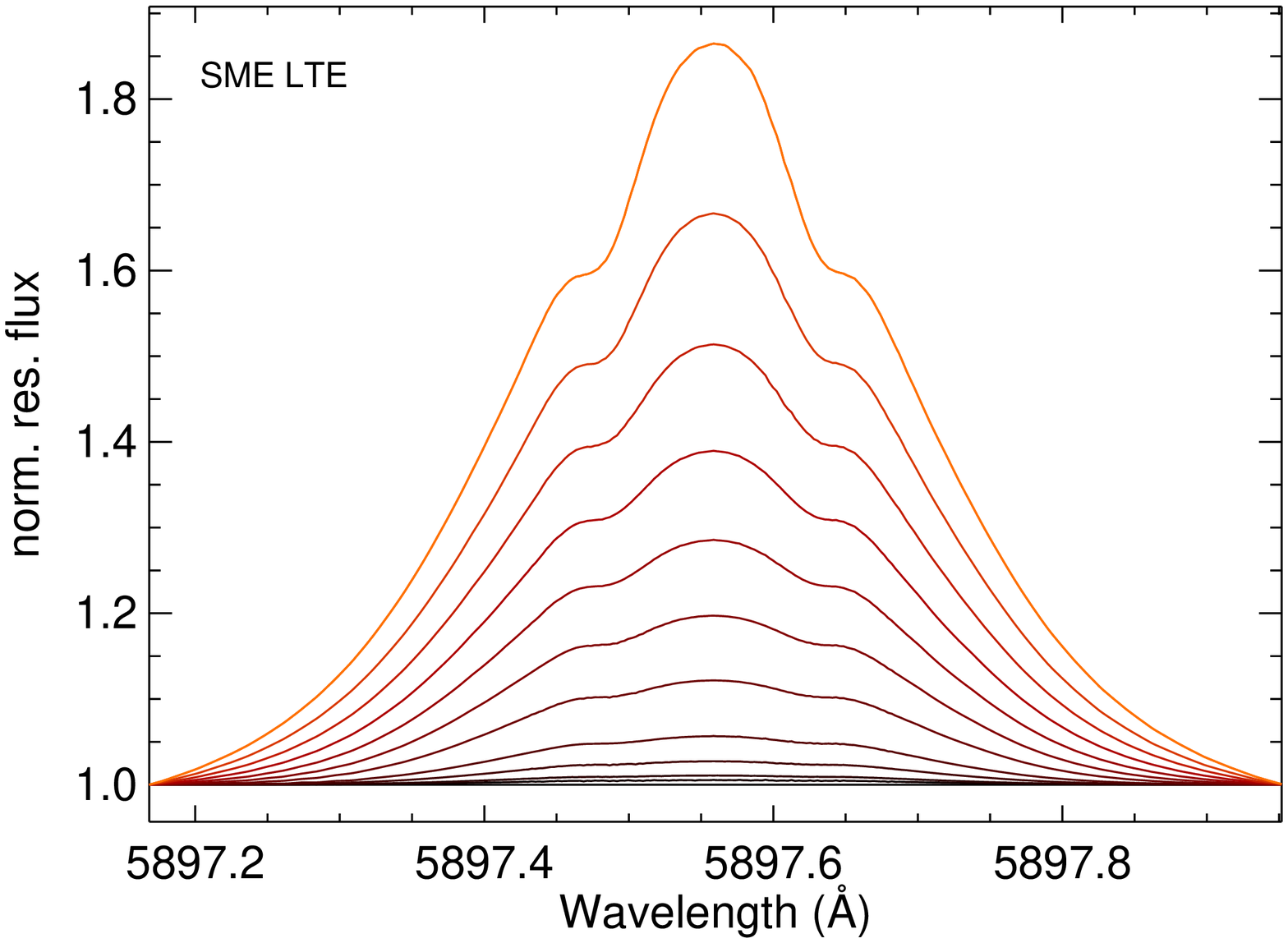}}
      \resizebox{.24\textwidth}{!}{\includegraphics[viewport=35 10 622 442]{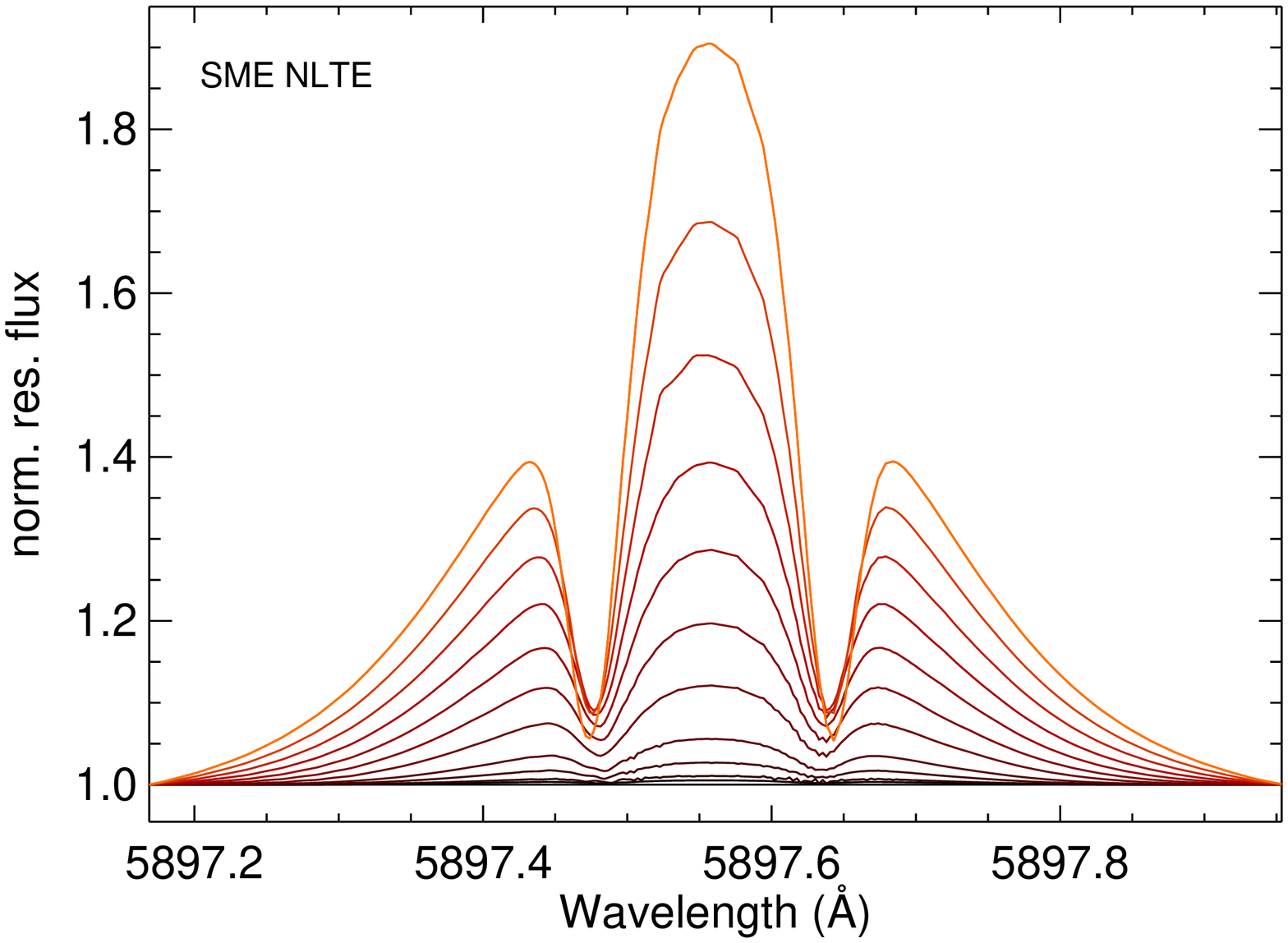}}
      \resizebox{.24\textwidth}{!}{\includegraphics[viewport=35 10 622 442]{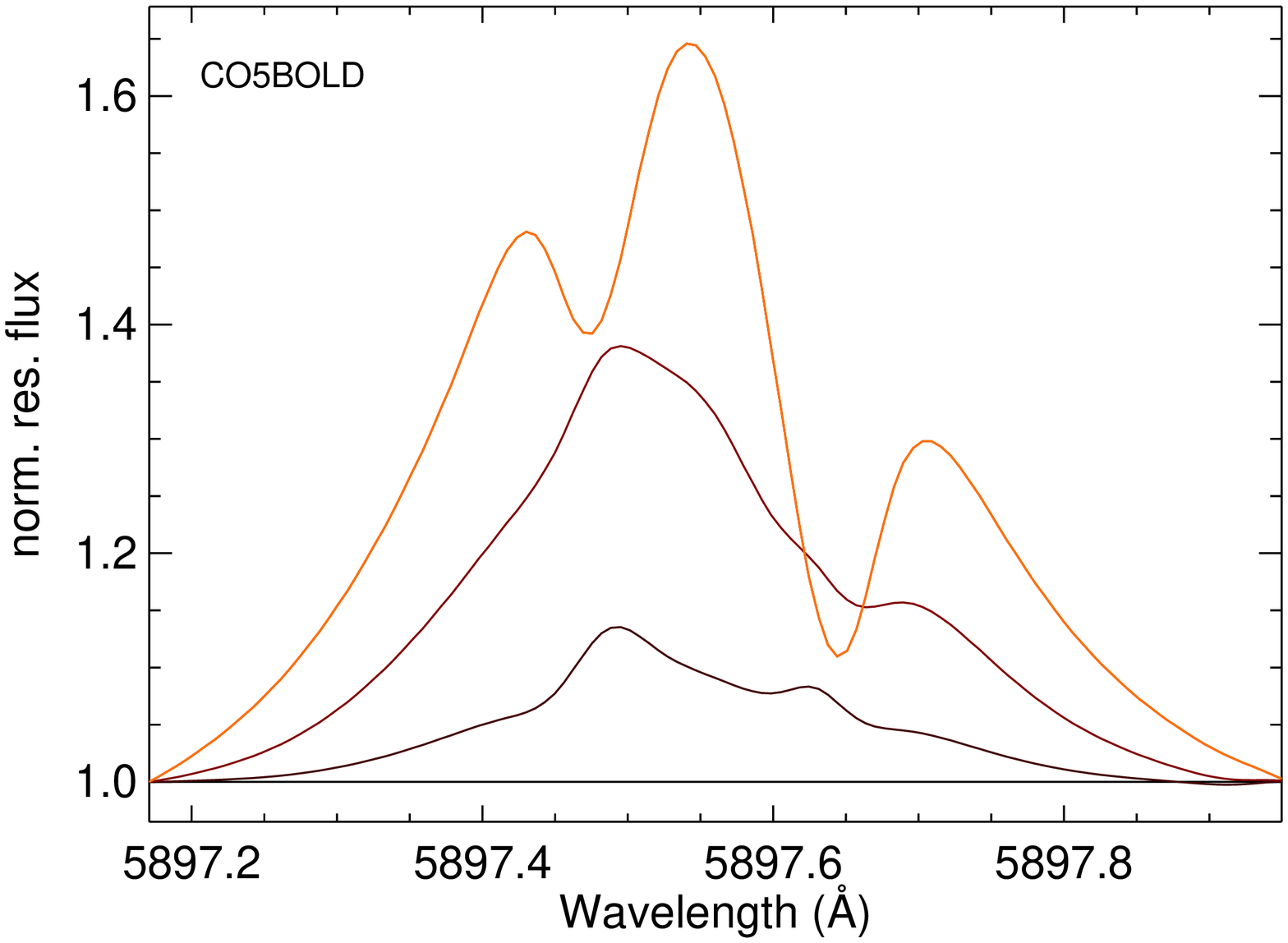}}
    }}
  \mbox{\parbox{.95\textwidth}{\tiny \vspace{3mm} \hspace{31mm} \ion{Mg}{i} \hspace{40mm} \ion{Mg}{i} \hspace{40mm} \ion{Mg}{i}\hspace{40mm} \ion{Mg}{i}}}\\[-5mm]
  \mbox{
    \parbox{\textwidth}{
      \resizebox{.24\textwidth}{!}{\includegraphics[viewport=35 10 622 442]{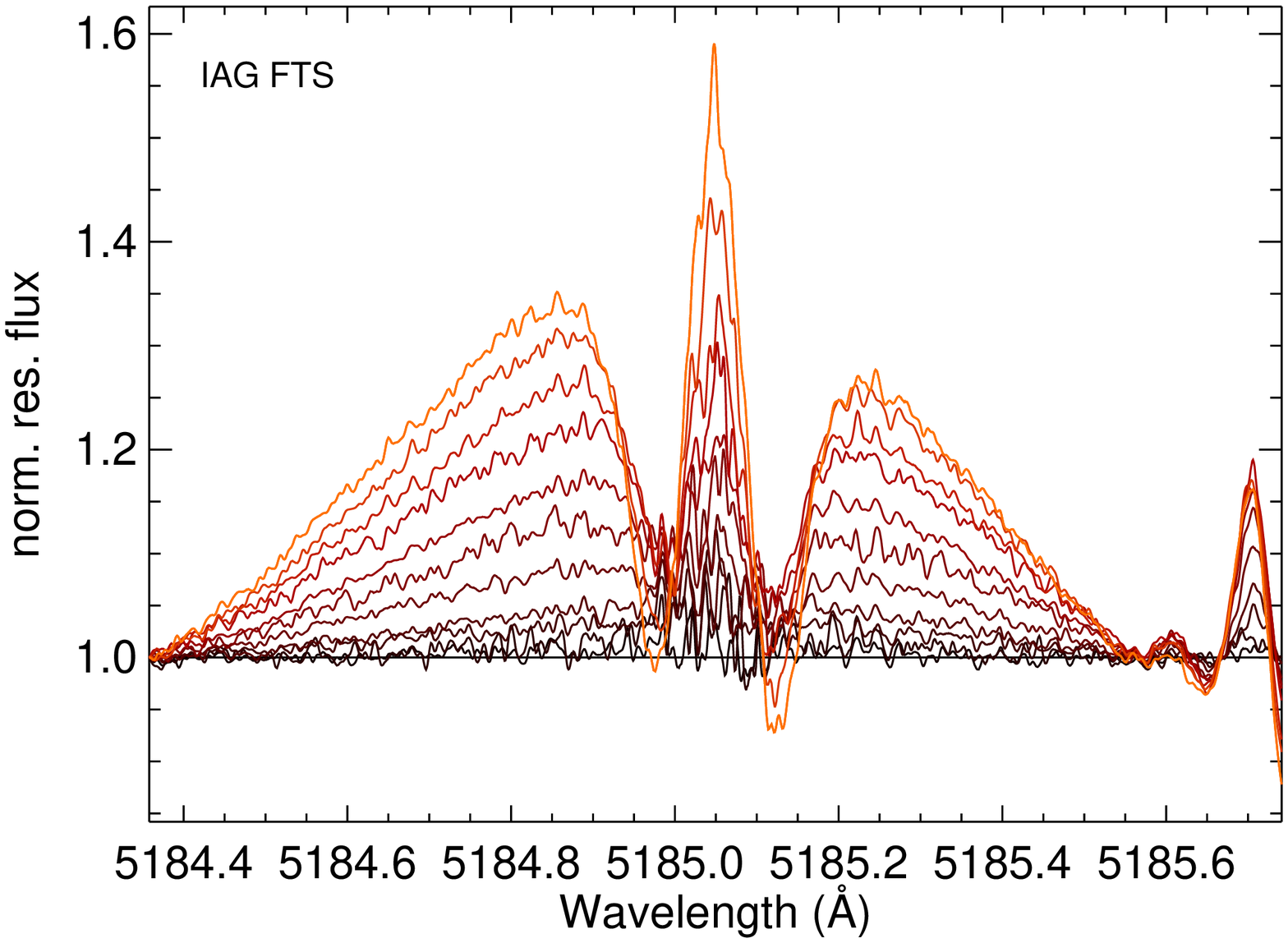}}
      \resizebox{.24\textwidth}{!}{\includegraphics[viewport=35 10 622 442]{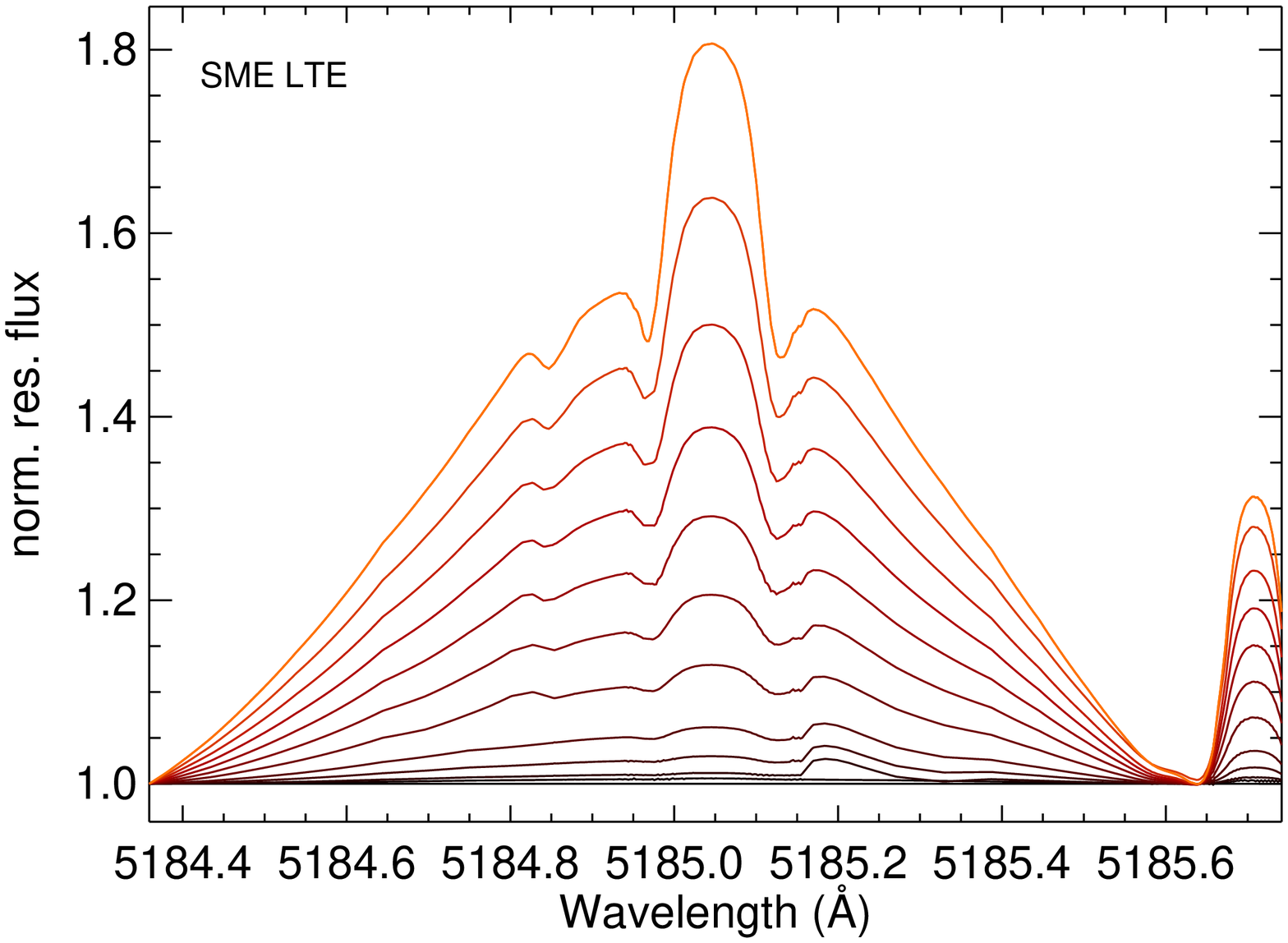}}
      \resizebox{.24\textwidth}{!}{\includegraphics[viewport=35 10 622 442]{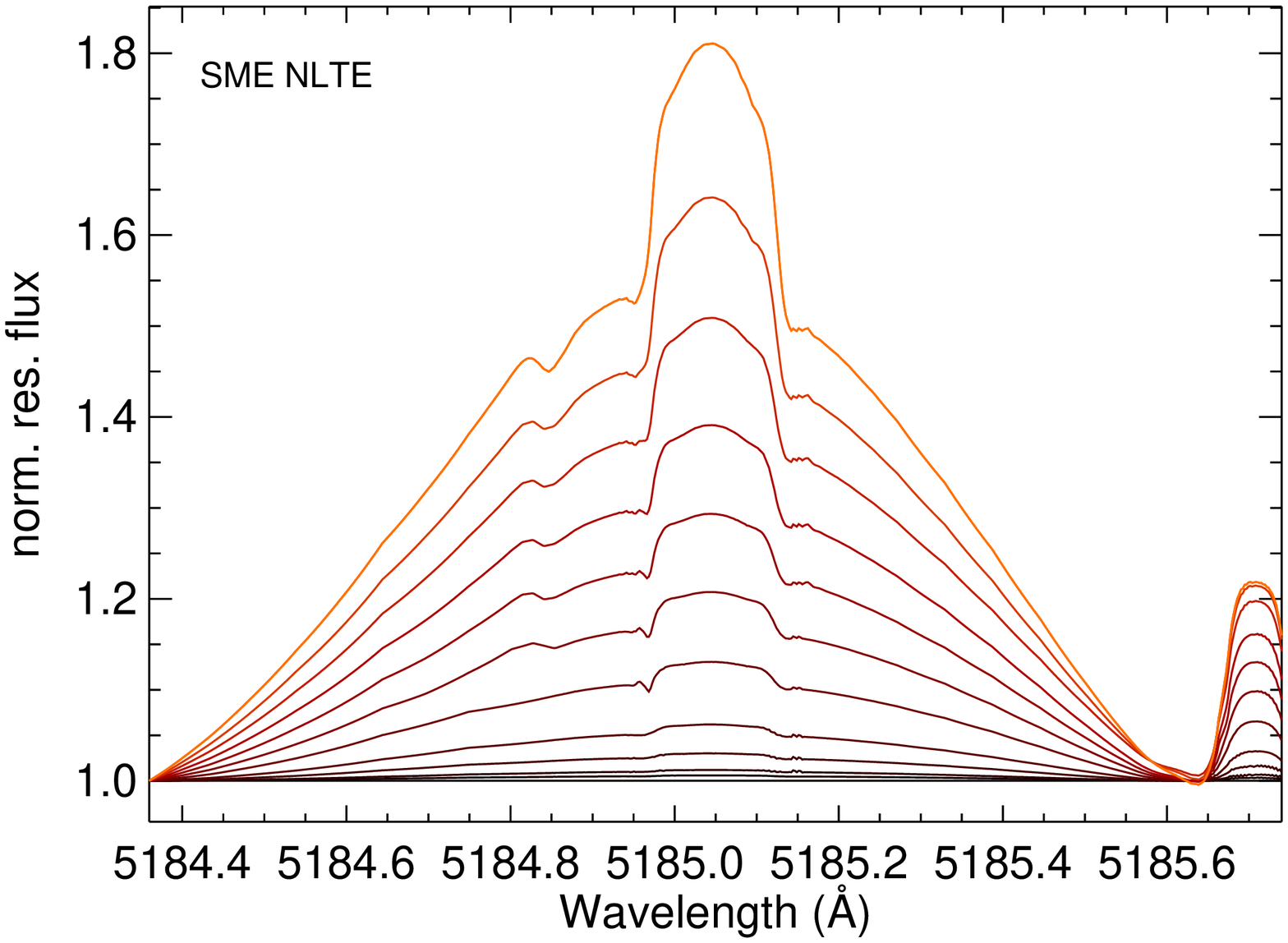}}
      \resizebox{.24\textwidth}{!}{\includegraphics[viewport=35 10 622 442]{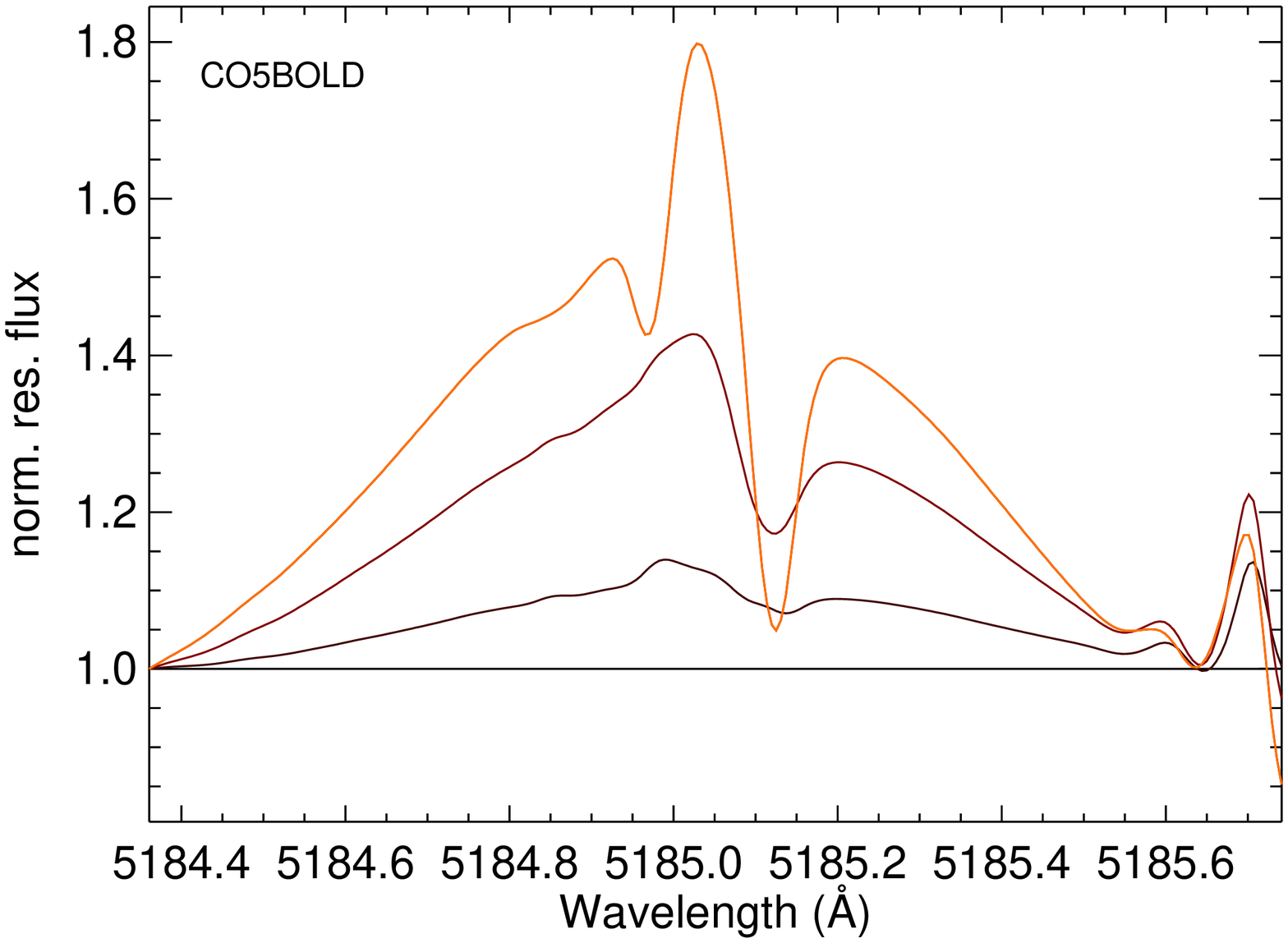}}
    }}
  \mbox{\parbox{.95\textwidth}{\tiny \vspace{3mm} \hspace{31mm} \ion{Ca}{i} \hspace{40mm} \ion{Ca}{i} \hspace{40mm} \ion{Ca}{i}\hspace{40mm} \ion{Ca}{i}}}\\[-5mm]
  \mbox{
    \parbox{\textwidth}{
      \resizebox{.24\textwidth}{!}{\includegraphics[viewport=35 10 622 442]{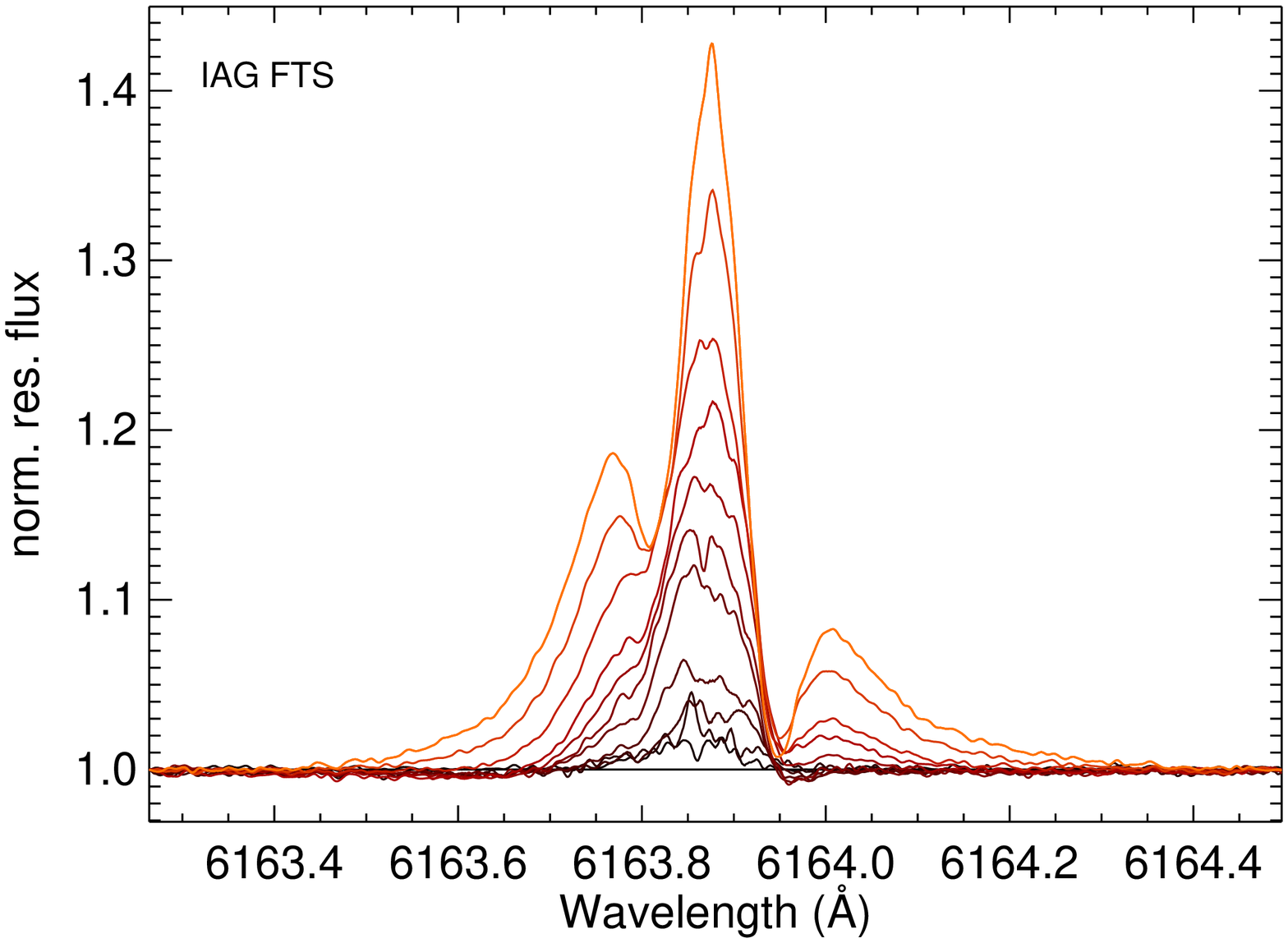}}
      \resizebox{.24\textwidth}{!}{\includegraphics[viewport=35 10 622 442]{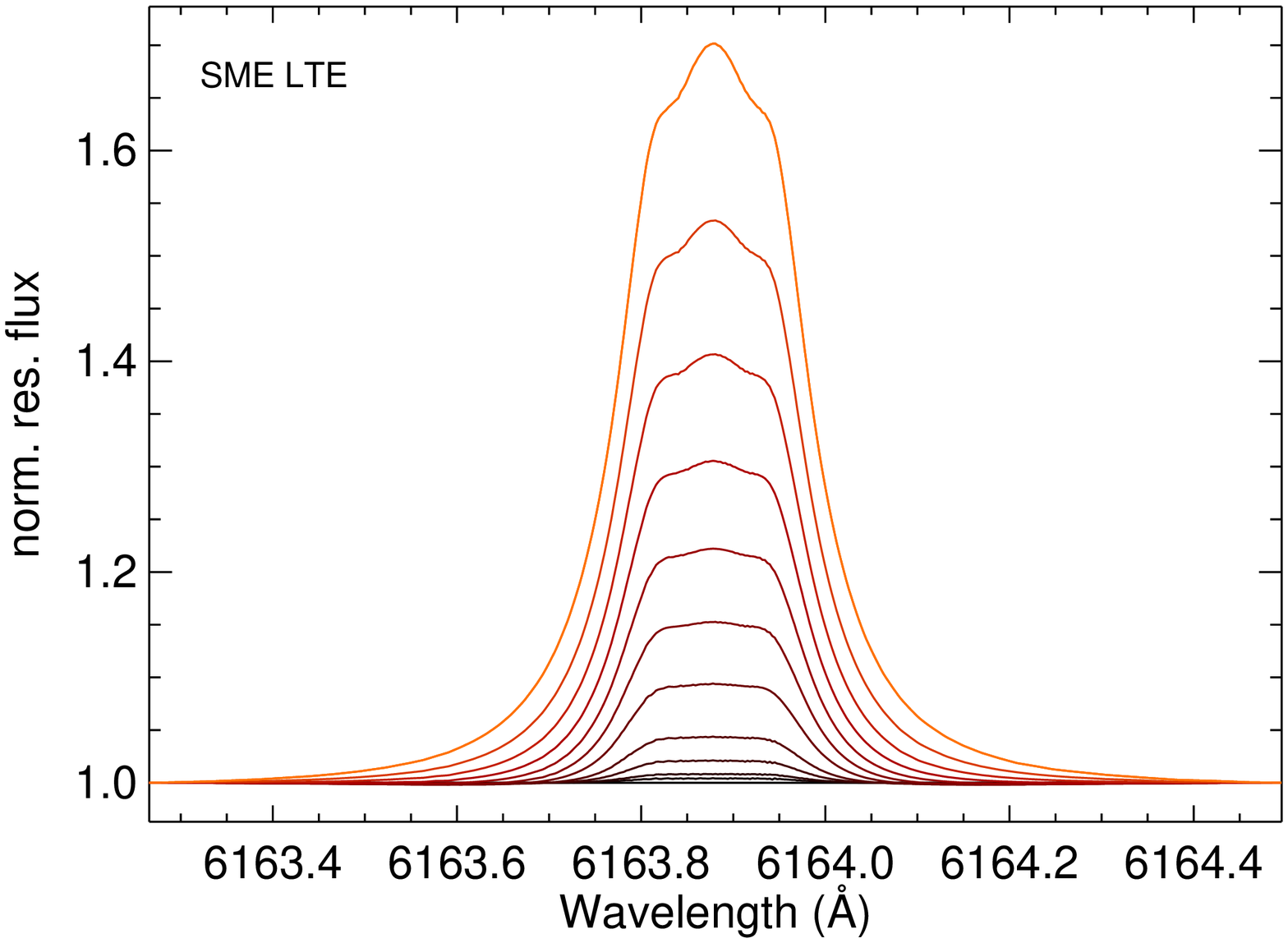}}
      \resizebox{.24\textwidth}{!}{\includegraphics[viewport=35 10 622 442]{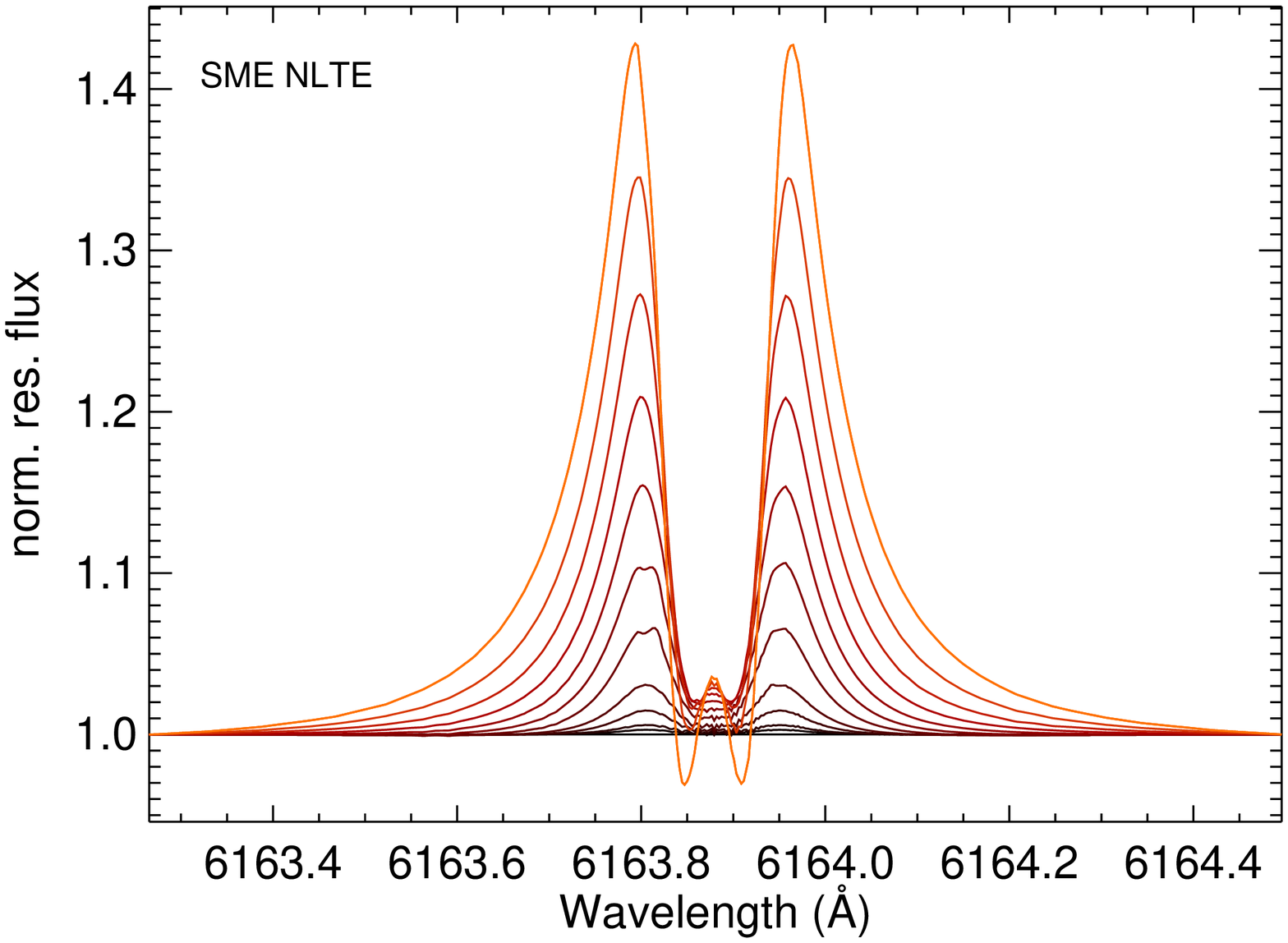}}
      \resizebox{.24\textwidth}{!}{\includegraphics[viewport=35 10 622 442]{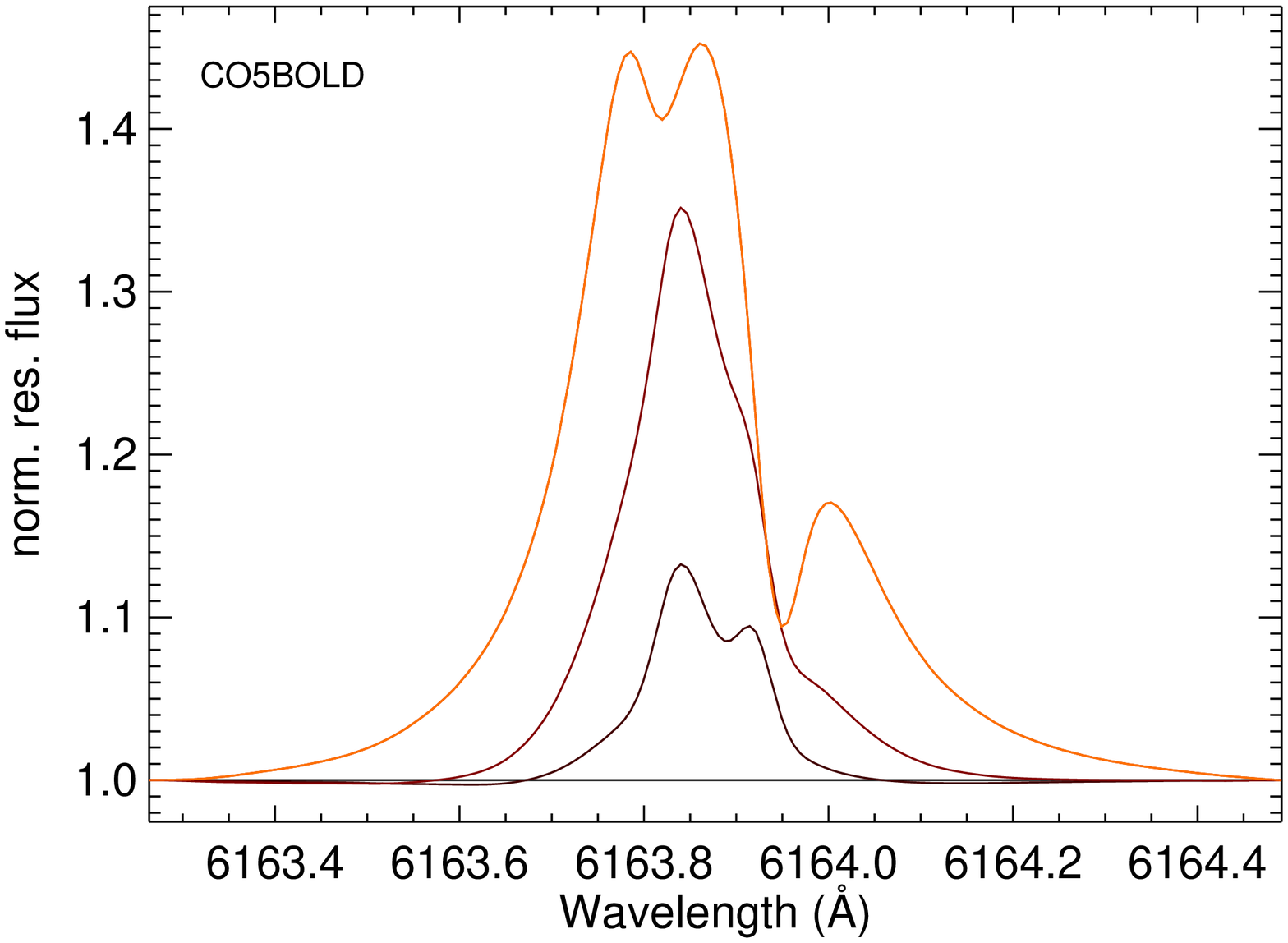}}
    }}
  \mbox{\parbox{.95\textwidth}{\tiny \vspace{3mm} \hspace{31mm} \ion{Fe}{i} \hspace{40mm} \ion{Fe}{i} \hspace{40mm} \ion{Fe}{i}\hspace{40mm} \ion{Fe}{i}}}\\[-5mm]
  \mbox{
    \parbox{\textwidth}{
      \resizebox{.24\textwidth}{!}{\includegraphics[viewport=35 10 622 442]{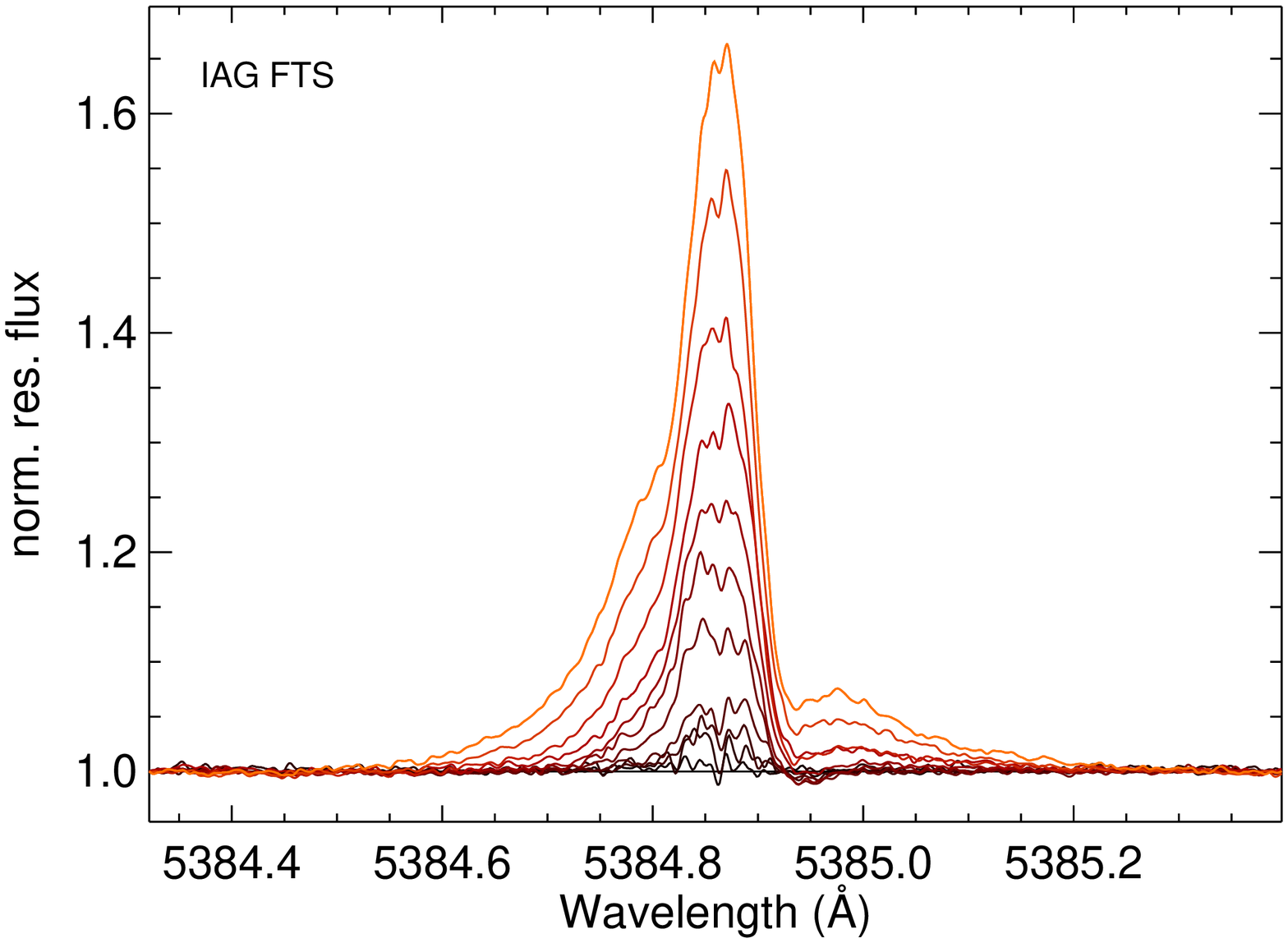}}
      \resizebox{.24\textwidth}{!}{\includegraphics[viewport=35 10 622 442]{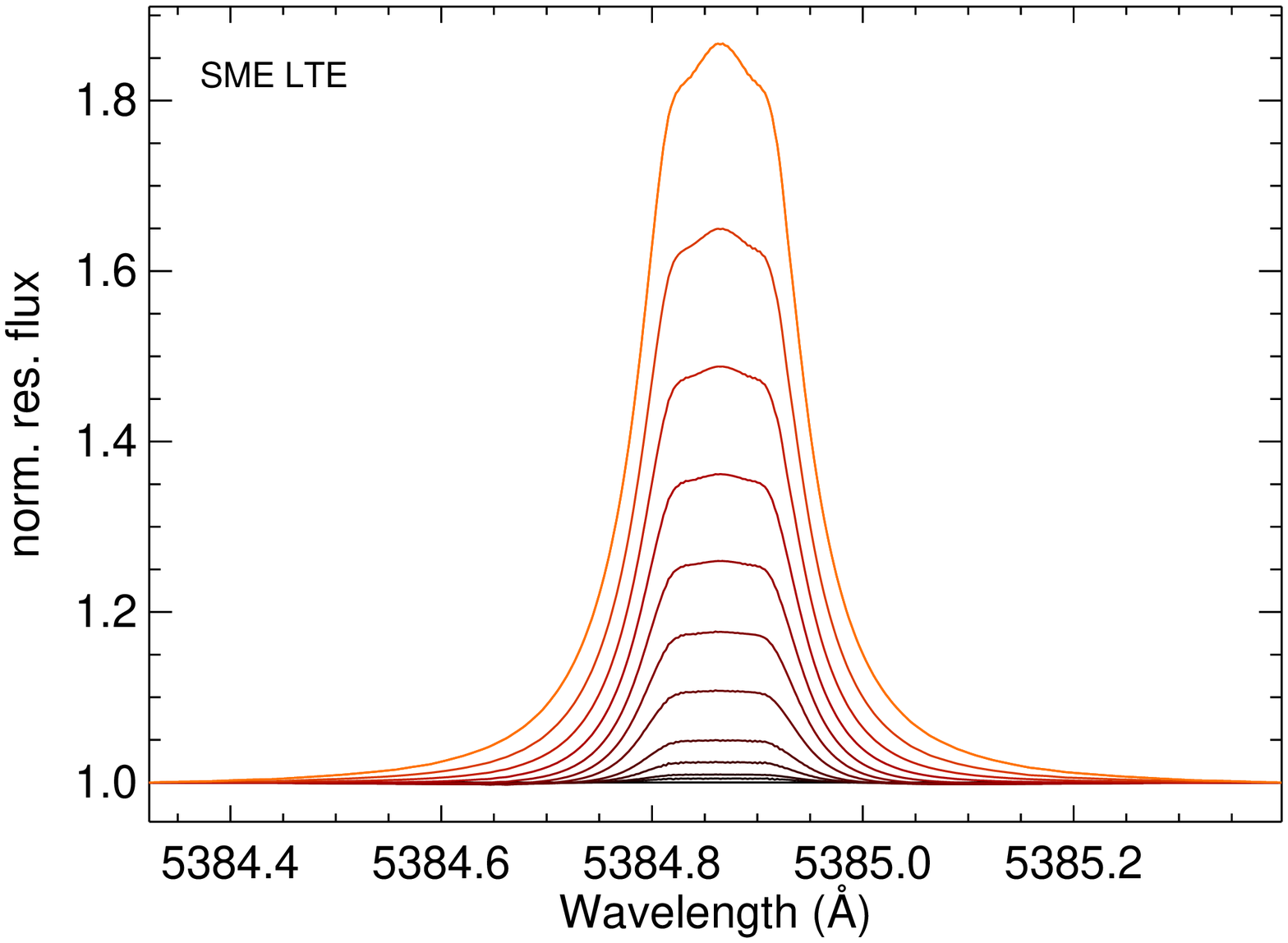}}
      \resizebox{.24\textwidth}{!}{\includegraphics[viewport=35 10 622 442]{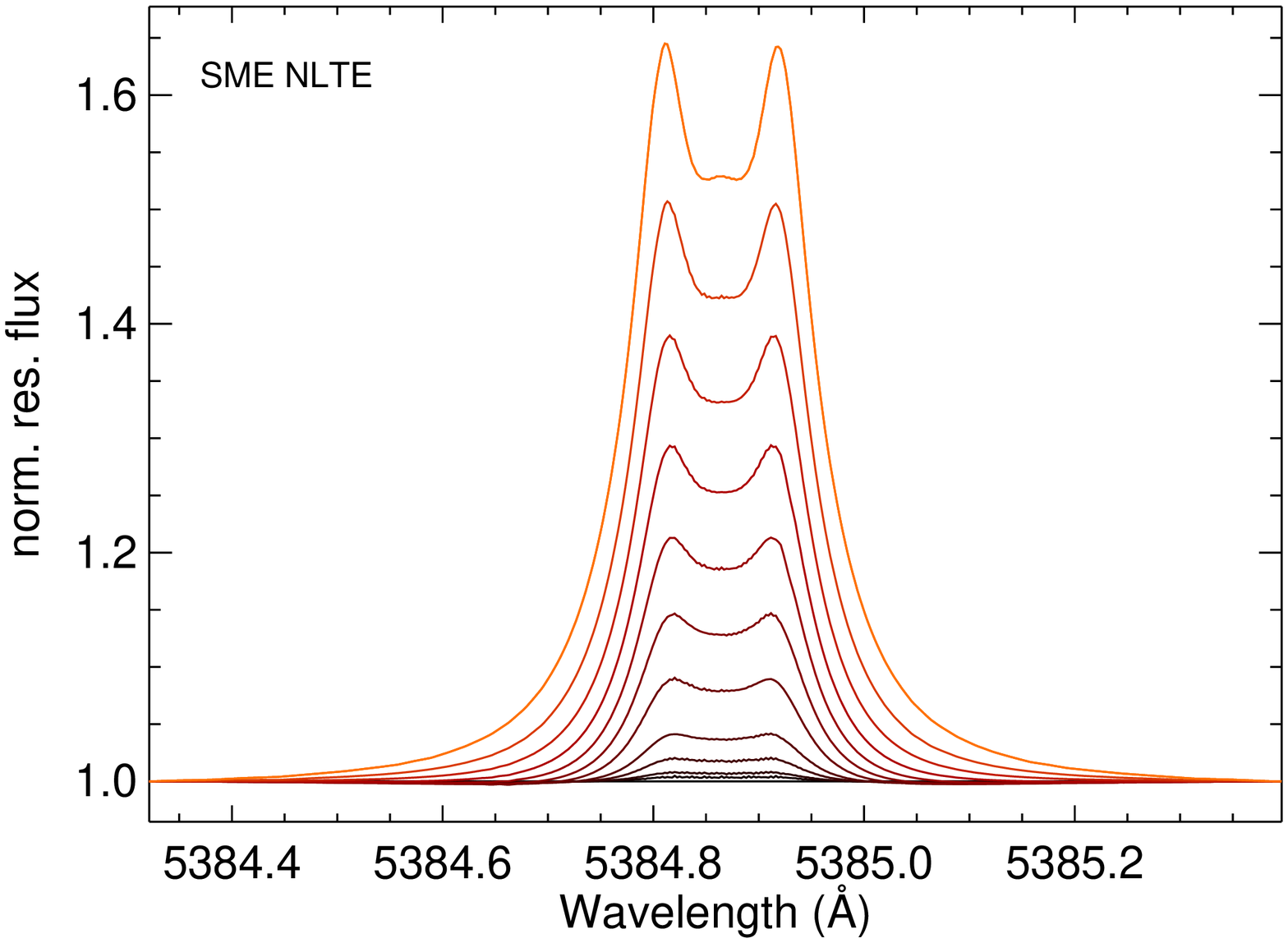}}
      \resizebox{.24\textwidth}{!}{\includegraphics[viewport=35 10 622 442]{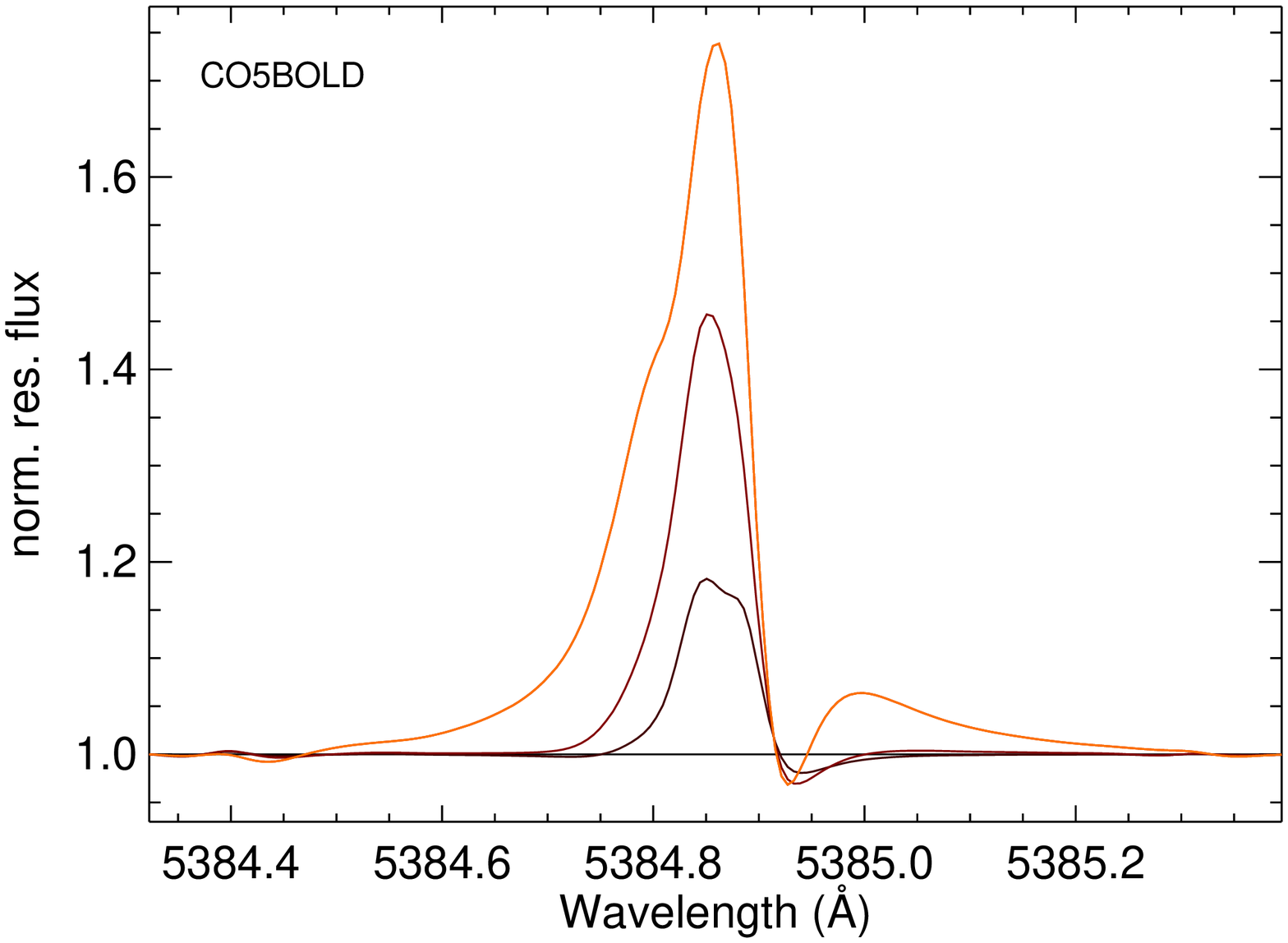}}
    }}
  \caption{\label{fig:LineRatios}Line ratios between spectra at different limb
    positions and the spectrum taken at disk center for four example lines:
    \ion{Na}{I} 5897, \ion{Mg}{I} 5185, \ion{Ca}{I} 6163, and \ion{Fe}{I} 5384
    (from top to bottom, panels span velocity ranges $\pm20, \pm40, \pm30$, and
    $\pm30$\,km\,s$^{-1}$, respectively).  The panels from left to right show observed IAG FTS spectra,
    and spectra for SME~LTE, SME~NLTE, and \cobold calculations
    as dashed lines, respectively. For IAG FTS and SME spectra, limb positions
    $\mu = 1.0, 0.99, 0.98, 0.95, 0.9, 0.8, 0.7, 0.6, 0.5, 0.4, 0.3, 0.2$ are
    shown. The \cobold models are available for $\mu = 1.0, 0.79, 0.41,
    0.09$. A darker color is used for spectra taken closer to disk center.}
}
\end{figure*}

The full spectral library and all model spectra are available online. Users
can inspect CLV in all spectral lines contained in the wavelength range and
compare the three different models to observations. Here, we show four lines
as typical examples of CLV and discuss the influence of NLTE and 3D convection
on the spectral lines, and compare them to our observations. In
Fig.\,\ref{fig:CLV}, we show spectral lines of \ion{Na}{i}, \ion{Mg}{i},
\ion{Ca}{i}, and \ion{Fe}{i} from top to bottom. Each panel shows the observed
FTS data for five limb positions $\mu = 0.2, 0.4, 0.6, 0.8$, and 1.0. From
left to right, we overplotted the three sets of model spectra, SME~LTE,
SME~NLTE, and \cobold (LTE and 3D convection). In Fig.\,\ref{fig:LineRatios},
we show the ratio between spectra from different limb positions and spectra at
disk center for the observations and models for the same four lines, that is,
in this figure the lines for $\mu=1$ are constant at ratio 1. We discuss
trends for the observations and models in the following.

\subsection{IAG FTS observations}

The observed IAG FTS data are shown as solid
lines in Fig.\,\ref{fig:CLV}, values of $\mu$ are color-coded, and black lines
show $\mu = 1$. The spectra for individual limb positions are averages
observed across the solar disk that are corrected for solar rotation. The
spectral lines' maximum absorption exceeds 80\,\% at line center in all four
lines. The left panel in Fig.\,\ref{fig:LineRatios} shows the ratios between
IAG FTS observations with respect to disk center.

The \textbf{\ion{Na}{i} line} (top panel in Figs.\,\ref{fig:CLV} and
\ref{fig:LineRatios}) shows a relatively broad line center and steep flanks
around the core. Damping wings are less steep. The line core remains
relatively constant for different $\mu$ values, and the damping wings show a
discernable CLV of approximately 5\,\% absorption depth between $\mu$ values
with $\Delta\mu = 0.2$. Ratios in Fig.\,\ref{fig:LineRatios} show that the
line asymmetry shifts from the disk center to limb. We note that we use the
\ion{Na}{i} line centered at $\lambda = 5897.6$\,\AA. We show the line only in
a range $\Delta \varv = \pm 20$\,km\,s$^{-1}$ around the line center. This is
because of a telluric line that is located in the right wing of the line. The
\ion{Na}{i} at 5891.6\,\AA\ (not shown in this work) is even more severely
contaminated by telluric lines. For the \ion{Mg}{i} line, we show the range
within $\Delta \varv = \pm 40$\,km\,s$^{-1}$, and for \ion{Ca}{i} and
\ion{Fe}{i}, we show $\Delta \varv = \pm 30$\,km\,s$^{-1}$.  For this work, we
did not attempt any telluric correction of the IAG FTS spectra. The wavelength
ranges selected show little impact of telluric contamination such that a
correction would not significantly alter the results.

The \textbf{\ion{Mg}{i} line} (second panel) is similar in shape to the
\ion{Na}{i} line, but the center is more curved and the damping wings are
stronger. The CLV follows a similar trend with little variation in the core
but significantly weaker absorption closer to the solar limb in the wings. The
latter is more clearly visible than in the \ion{Na}{i} lines because the
damping wings appear less steep and lines from different limb positions appear
more clearly separated. The line symmetry is relatively constant across the
disk, which can be seen in the line ratios in Fig.\,\ref{fig:LineRatios}.

The \textbf{\ion{Ca}{i}} line (third panel) and also the \textbf{\ion{Fe}{i}}
line (bottom panel) show smooth Voigt profiles. The CLV in both lines is
relatively weak. Lines closer to the limb appear a few percent weaker than
lines at disk center, but the overall shape remains largely unchanged. Line
ratios in Fig.\,\ref{fig:LineRatios} reveal the shift of the line core from
disk center to limb.

\subsection{SME LTE}

We are comparing model line profiles from SME LTE calculations to our
observations in the left panel of Fig.\,\ref{fig:CLV}. In general, the line
cores in the calculations match the cores of the IAG FTS data relatively
well. Toward the wings, the difference between the model and observations varies
across the limb. This could be explained by convective blueshift, which is not
taken into account in this model. All line cores of the model appear
significantly weaker at the solar limb with respect to disk center. None of
the observed lines shows CLV as strong as predicted in the SME LTE model.

The line ratio plots in the second panel of Fig.\,\ref{fig:LineRatios} reflect
the overall shape of lines and the way intensity changes differently at
different line depths. All patterns show the same symmetry for each line
because radiative transfer in the absence of a velocity field linked with
convective motions does not produce any asymmetries \citep{Asplund2009}.

\subsection{SME NLTE}

The line profiles from the SME NLTE calculations differ significantly from the
LTE calculations in all four lines.  In the \ion{Na}{i} line, the NLTE
calculations provide a reasonable match of line absorptions and CLV to the
observations. This is opposite to the \ion{Mg}{i} line for which the NLTE model
predicts weaker line cores than the LTE model (but strong CLV), which means
the mismatch becomes even larger. For the \ion{Ca}{i} line, the NLTE model
shows significantly weaker CLV than the LTE model. Otherwise, the difference
between LTE and NLTE is relatively little in the \ion{Ca}{i} and also in the
\ion{Fe}{i} line. As for the SME LTE case, the SME NLTE line cores match the
position of the observations relatively well.

A comparison between LTE and NLTE calculations shows that among the four lines
we selected, the \ion{Na}{i} line is most sensitive to NLTE effects. In this
line, the line depths in the LTE calculations are substantially smaller than
in the observations, and the line depths are significantly larger in the NLTE
calculations. A similar situation can be observed in the \ion{Ca}{i} line;
although, it is not as obvious as in \ion{Na}{i}. In the other two lines, NLTE
effects are less pronounced.

\subsection{\cobold}
\label{sect:coboldcalc}

The \cobold calculations were carried out in LTE. In general, the CLV of line
strengths is similar to the SME LTE model; although, the amplitude of the
variability in line strength at the core is diminished and closer to the
observed behavior than SME LTE. In all four lines, the inclusion of convection
in the model calculations provides some improvement against the
1D model (SME LTE). The residuals between observations and the
\cobold spectra are particularly small for the \ion{Fe}{i} line, while rather
significant differences appear in the other lines (Fig\,\ref{fig:CLV}). The
effect of convection is clearly visible in the asymmetry of the line profile
models, in particular in the blue line wings (see ratio panels in
Fig\,\ref{fig:CLV}). Including convective motion in the line formation
calculations introduces an asymmetry of the line profiles that shifts the line
core by several hundred m\,s$^{-1}$ redward from center to limb. The effect
is stronger than observed in the IAG FTS data.

In the right panel of Fig.\,\ref{fig:LineRatios}, we show the line ratios from
the \cobold calculations. In this figure, we show only calculations for which
radiative transfer calculations were carried out (we avoided
interpolations). The line ratios in the right panel of
Fig.\,\ref{fig:LineRatios} show that the modeled profile asymmetry varies
across the solar disk. This is because the projection of convective velocities
toward the observer is not symmetric. It is interesting to compare the shape
of ratios between the \cobold calculations and the IAG FTS observations. The
change in line ratios predicted by the \cobold model shows some similarity to
the observed behavior, which can be interpreted as the modeled convection
partially resembling the convective velocity field on the solar surface.

\section{Synthetic transit models}
\label{sect:transit}

\begin{figure}
  \centering
  \resizebox{0.8\hsize}{!}{\includegraphics{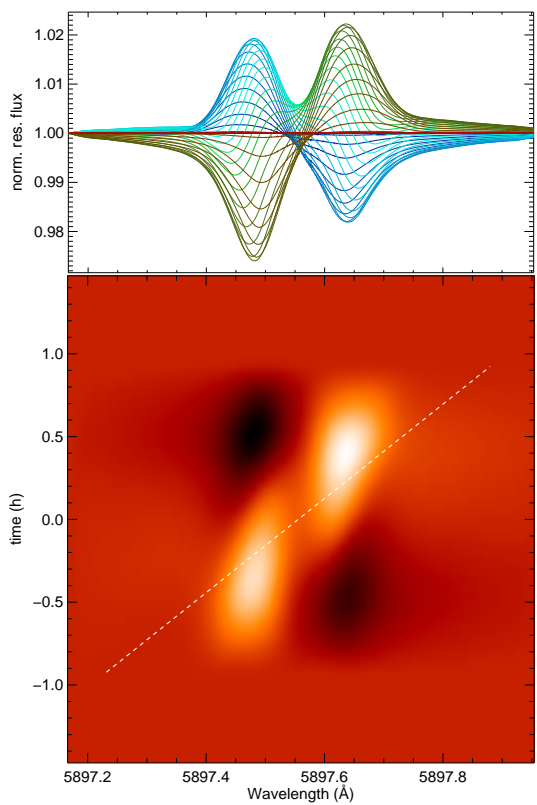}}
  \caption{\label{fig:NaDynSpec}Spectroscopic signature of the transiting
    planet in the \ion{Na}{i} line. \emph{Top panel:} Residuals between
    spectra during the transit and out-of transit spectrum, with different colors
    indicating the time during transit. \emph{Lower panel:} Residuals shown as
a     dynamic spectrum with color indicating normalized residual flux, and a brighter
    color indicating larger values. The white dashed line shows the planetary
    path. The plot covers a velocity range of $\pm20$\,km\,s$^{-1}$ around the
    line center.}
\end{figure}

We used the FTS IAG atlas to simulate planetary transit observations including
the effect of CLV. Furthermore, we compared our results to simulations from
radiative transfer calculations with SME LTE, SME NLTE, and \cobold as in
earlier sections, with the goal to estimate the systematics that can be
expected if CLV is not taken into account. Our simulation carried out
numerical integration over the visible surface of the stellar disk. The
stellar surface was approximated by a grid with elements equal in size and 201
elements along the equator, which provides sufficient spatial resolution for
our simulations. During transit, the elements occulted by the planet were
neglected in the integration. For each surface element, we interpolated the
local spectrum of the FTS IAG atlas according to the element's value of
$\mu$. For the limb darkening law, we chose the tabulation of
\citet{1994SoPh..153...91N} and selected the entry according to the wavelength
closest to the spectral line. For each element, the spectrum was shifted
according to its projected velocity toward the observer.

\begin{table}
  \centering
  \caption{Configuration of planetary orbit and stellar parameters used for
    the transit simulations are similar to the HD\,189733 system. Parameters
    are adopted from \citet{2009A&A...506..377T} and \citet{2017A&A...603A..73Y}.}
  \label{tab:configuration}
  \begin{tabular}{lr}
    \hline\hline\noalign{\smallskip}
    Parameter  & Value \\
    \hline\noalign{\smallskip}
    Planet mass (M$_\textrm{Jup}$)  & 1.138\\
    Planet radius (R$_\textrm{Jup}$) & 1.505\\ 
    Orbital period (d) & 2.218573\\
    Star radius (R$_{\odot}$) & 0.756\\
    Star mass (M$_{\odot}$) & 0.806\\
    Star $P_\textrm{rot}$ (d) & 12.41\\
    Inclination of stellar rotation axis (deg) & 87\\
    Star $\varv \sin{i}$ (km\,s$^{-1}$) & 3.08\\
    Eccentricity & 0\\
    Orbital inclination (deg) & 85.508\\
    Spin-orbit angle $\lambda$ (deg) & $-0.85$\\
    Impact parameter $a$ & 0.69\\
    \hline\noalign{\smallskip}
  \end{tabular}
\end{table}

We demonstrate the effect of CLV for the example of a planet in orbit around a
star. We use solar spectra from the IAG FTS atlas and from solar spectrum
models. For our examples, we assume a star-planet configuration similar to
HD\,189733b, that is, geometry and parameters of the planetary orbit resemble
the HD~189733 system (see Table\,\ref{tab:configuration}), but we use solar
spectra and neither observations nor atmosphere models of HD~189733 enter our
simulation. We emphasize that HD~189733 is cooler than the Sun and our local
spectra are therefore not representative for this particular
star. Nevertheless, our simulations provide an estimate for the order of
magnitude of systematic effects that can be expected in a system similar to
HD~189733.

We show an example dynamic spectrum for the \ion{Na}{i} line in
Fig.\,\ref{fig:NaDynSpec}. The lower panel visualizes the residuals between
the average out-of-transit spectrum of the star and the spectral line during
transit in the range $\pm 20$\,km\,s$^{-1}$ around line center. The radial
velocity path of the planet during transit is overplotted as a dashed white
line. The planetary path is similar to the spectral transit feature (the RM
signature) if the planetary orbit and stellar rotation are aligned
($\lambda = 0$), and they also coincide if
$P_{\textrm{orb}} = P_{\textrm{rot}}$. For example, the planetary path closely
follows the transit feature in HD~209458 and precise correction of the RM
feature is particularly important \citep[see][]{2021A&A...647A..26C}. In the
following, transmission curves are calculated in the range
$\pm20$\,km\,s$^{-1}$ around the line center, that is, the same range as shown in
Fig.\,\ref{fig:NaDynSpec}.

\subsection{Transmission curves}
\label{sect:transmissioncurve}

\begin{figure}
  {\sffamily
  \centering
  \mbox{\parbox{.95\textwidth}{\small \vspace{3mm} \hspace{70mm} \ion{Na}{i}}}\\[-5mm]
  \includegraphics[width=.9\hsize, viewport=8 5 625 450]{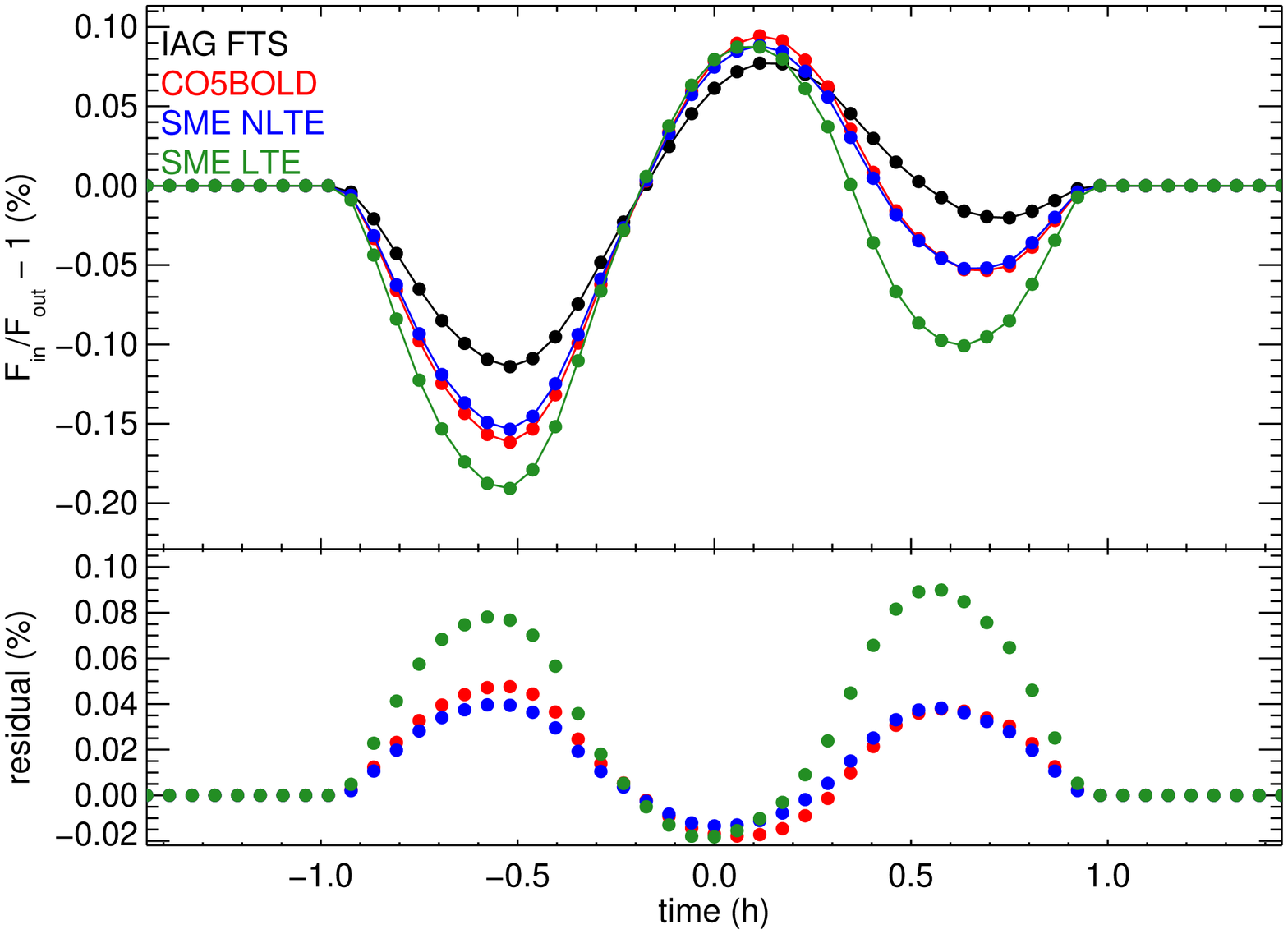}
  \mbox{\parbox{.95\textwidth}{\small \hspace{70mm} \ion{Mg}{i}}}\\[-5mm]
  \includegraphics[width=.9\hsize, viewport=8 5 625 450]{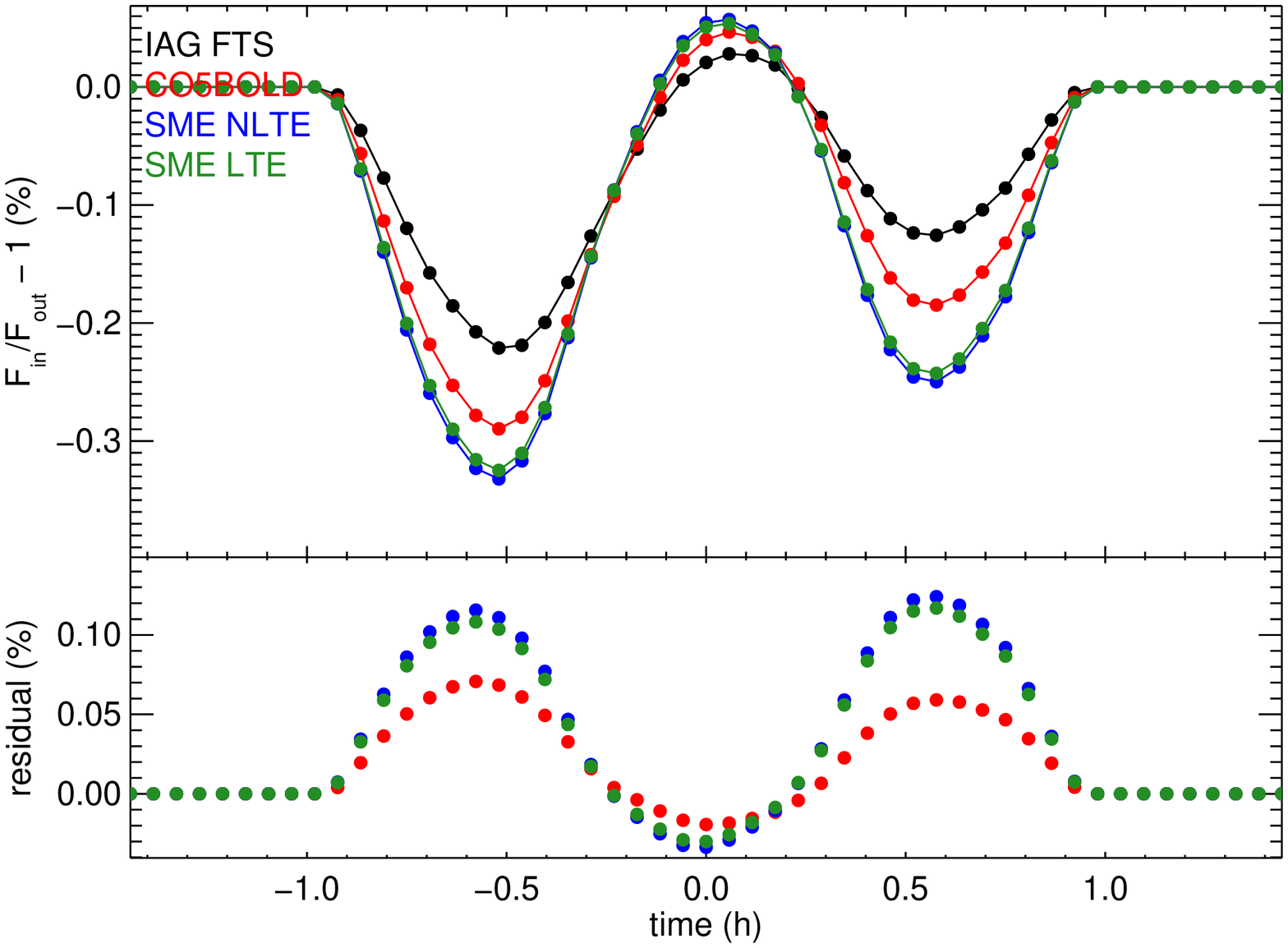}
  \mbox{\parbox{.95\textwidth}{\small \hspace{70mm} \ion{Ca}{i}}}\\[-5mm]
  \includegraphics[width=.9\hsize, viewport=8 5 625 450]{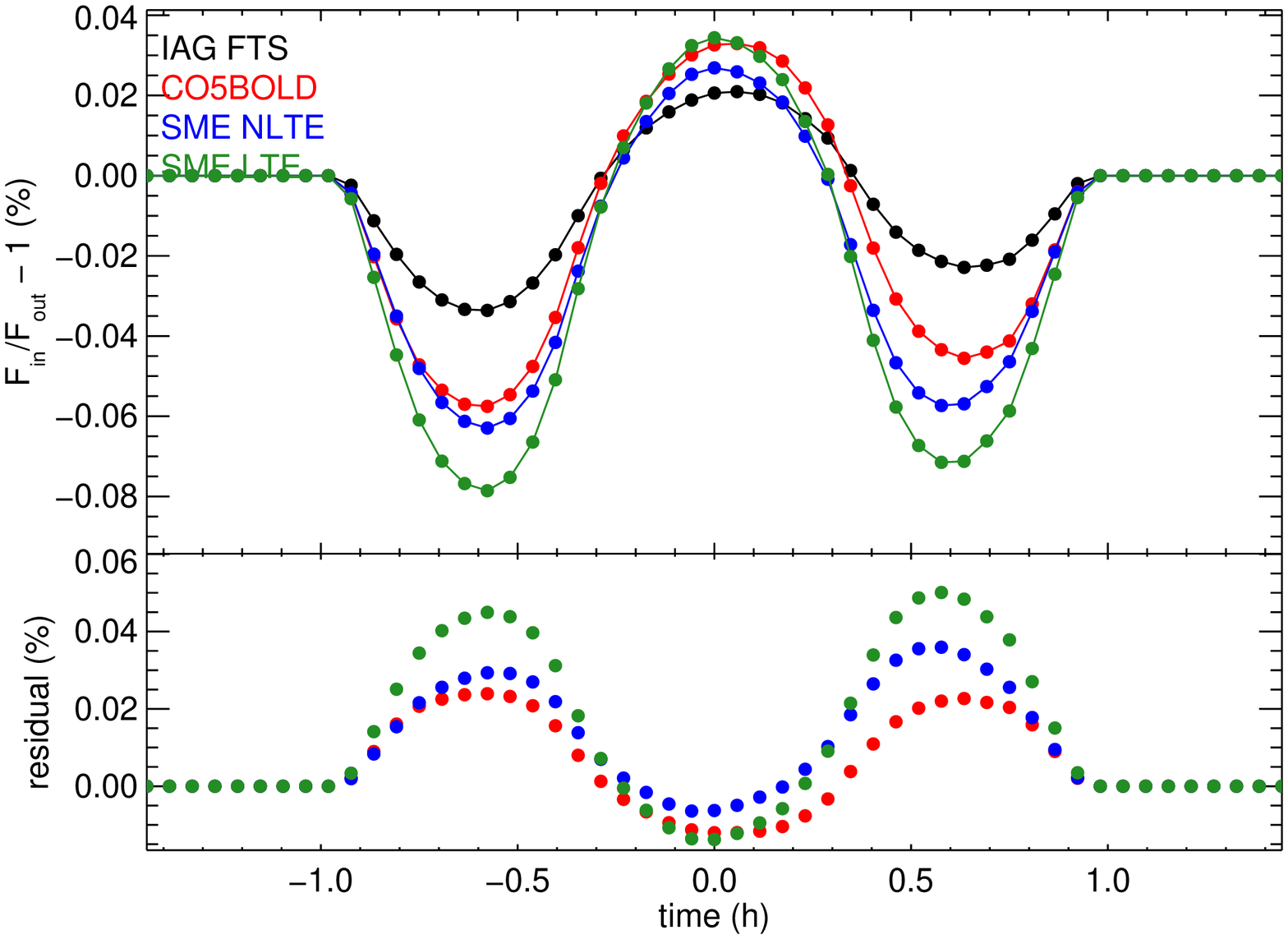}
  \mbox{\parbox{.95\textwidth}{\small \hspace{70mm} \ion{Fe}{i}}}\\[-5mm]
  \includegraphics[width=.9\hsize, viewport=8 5 625 450]{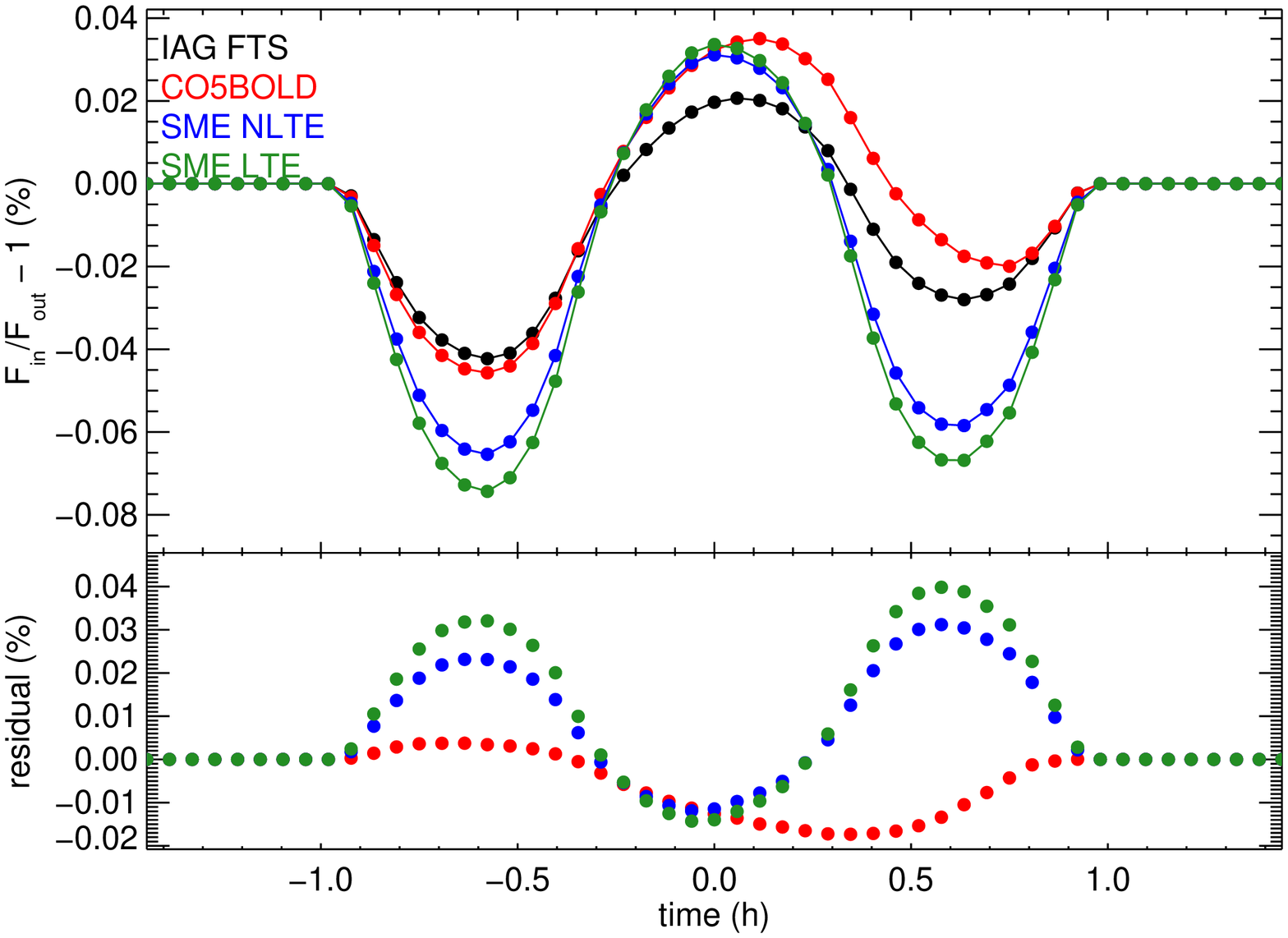}
  \caption{\label{fig:TM}Example transmission curves and residuals between IAG
    FTS and model transmission curves for \ion{Na}{i}, \ion{Mg}{i},
    \ion{Ca}{i}, and \ion{Fe}{i} (top to bottom). The configuration corresponds to
    the HD~189733b system (see Table\,\ref{tab:configuration}).}
  }
\end{figure}

We show transmission curves for the four example lines in
Fig.\,\ref{fig:TM}. The plots show the ratio between the normalized flux
observed ($F_\textrm{in}$) and the normalized flux out of transit
($F_\textrm{out}$). Fluxes are normalized because absolute flux information is
usually not retained in high-resolution spectroscopic observations at the
level required for this measurement. This means that the transmission curve,
$F_\textrm{in}/F_\textrm{out} - 1$, can be positive during transit even if the
total flux during transit is always smaller than when out of transit. The
transmission curve is negative if the spectral lines appear deeper during
transit, and it is positive if the spectral lines appear shallower. The
transmission curve can be computed from the dynamic spectrum
(Fig.\,\ref{fig:NaDynSpec}) by integration across the spectral domain (here,
$\pm 20$\,km\,s$^{-1}$).

The amplitude and shape of the transmission curves caused by the planetary
transit in absence of any atmospheric absorption depend on the choice of the
local line profiles. In all four lines, the effect is smallest if the observed
FTS IAG lines are used, and the amplitudes of the transmission curves are
overpredicted for all three model sets. This is a consequence of the CLV
effect being overestimated in the models as seen in Fig.\,\ref{fig:CLV}. In
general, the best match is typically found for the \cobold model and the SME
LTE model shows the largest deviation. The amplitude of the transmission curve
is overpredicted within a factor of 2--3, and the difference to the IAG FTS
results does not grow significantly larger than 1/1000.

The transmission curves are computed relative to the out-of-transit
spectrum. This means that for the influence of CLV on the transmission curve,
the relative differences between the line profiles observed at different limb
positions are relevant while absolute differences between line profile
calculations and the real profiles can be normalized out. The relative
differences between line profiles at different limb positions in the IAG FTS
\ion{Na}{i} lines (upper left panel in Fig.\,\ref{fig:LineRatios}) are more
closely resembled by the \cobold calculations and the SME NLTE calculations
than they are by the SME LTE calculations. This can explain why the \cobold
and SME NLTE transmission curves are closer to the IAG FTS transmission curve
than the SME LTE curve.

\begin{figure}
  \centering
  \resizebox{\hsize}{!}{\includegraphics[viewport=15 5 640 450]{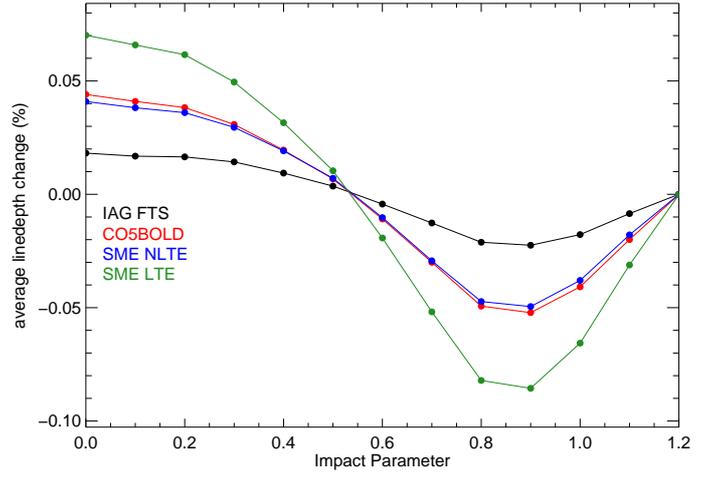}}
  \caption{\label{fig:ImpactStudyTM}Impact study showing the average line depth
    change for the \ion{Na}{i} line depending on the impact parameter,
    $a$. The configuration is the same as in Table\,\ref{tab:configuration}, but with orbital
    inclination according to the value of $a$.}
\end{figure}

Some of the transmission curves appear asymmetric. This is caused by
asymmetries in the line profiles and their selective occultation during
transit. Convective velocities can cause asymmetries in the observed IAG FTS
line profiles and in the \cobold\ calculations. Their transmission curves show
some asymmetry in all lines. For the SME calculations, no convection is taken
into account. Nevertheless, SME transmission curves are also significantly
asymmetric in the \ion{Na}{i} and \ion{Mg}{I} lines. This is because the line
profiles of \ion{Na}{i} and \ion{Mg}{I} show some asymmetry that is caused by
line blends, which is visible in the far wings of these lines (see
Fig.\,\ref{fig:CLV}). The \ion{Ca}{i} and \ion{Fe}{I} profiles, on the other
hand, are rather symmetric. We note that normalization of the local line
profiles can introduce systematic errors in the transmission curve, in
particular if the lines extend to the edges of the integration region as in
the \ion{Na}{i} and \ion{Mg}{i} examples. This can be relevant in transit
observations because the continuum level may be difficult to determine.

We carry out an analysis of the difference between the transmission curve for
the FTS IAG line profiles and the three model sets for a range of impact
parameters, $a$. The impact parameter describes the minimum distance between
the center of the star and the center of the planet during transit in units of
the stellar radius; a transit occurs as long as
$a < 1 + r_{\textrm{planet}}/r_{\textrm{star}}$ (grazing transits for
$a > 1$).  In Fig.\,\ref{fig:ImpactStudyTM}, we show the average line depth
change for impact parameters, $a$, between 0 and 1.2 for the case of the
\ion{Na}{i} line. We define the average line depth change as the integral of
the transmission curve (top panel in Fig.\,\,\ref{fig:TM}) divided by the
number of observations during transit. For impact parameters larger than
$a \sim 0.55$, the average line depth is negative while it is positive for
transits observed closer to the stellar center. This is because the
\ion{Na}{i} lines are deeper at disk center for all line profile sets (top
panel in Fig.\,\ref{fig:CLV}). Consistent with our result for the HD~189733b
configuration ($a = 0.69$), the SME LTE profiles show the largest effect for
all impact parameters. We conclude from Fig.\,\ref{fig:ImpactStudyTM} that the
transmission curves computed from the three models scale with the transmission
curve based on the FTS IAG profiles for all impact parameters.

\subsection{Rossiter-McLaughlin curves}

\begin{figure}
  {\sffamily
  \centering
  \mbox{\parbox{.95\textwidth}{\small \vspace{3mm} \hspace{70mm} \ion{Na}{i}}}\\[-5mm]
  \includegraphics[width=.9\hsize, viewport=8 5 625 450]{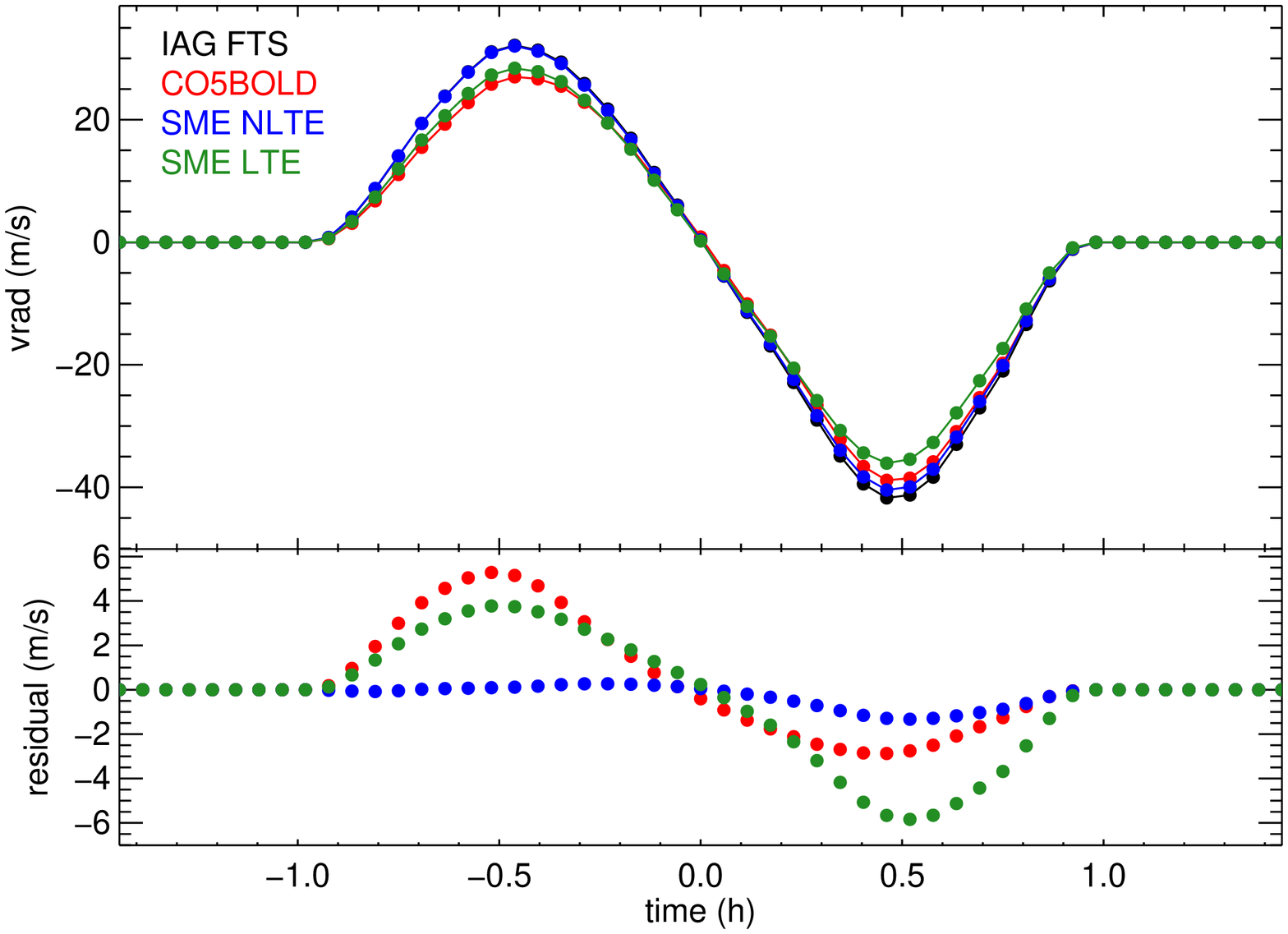}
  \mbox{\parbox{.95\textwidth}{\small \hspace{70mm} \ion{Mg}{i}}}\\[-5mm]
  \includegraphics[width=.9\hsize, viewport=8 5 625 450]{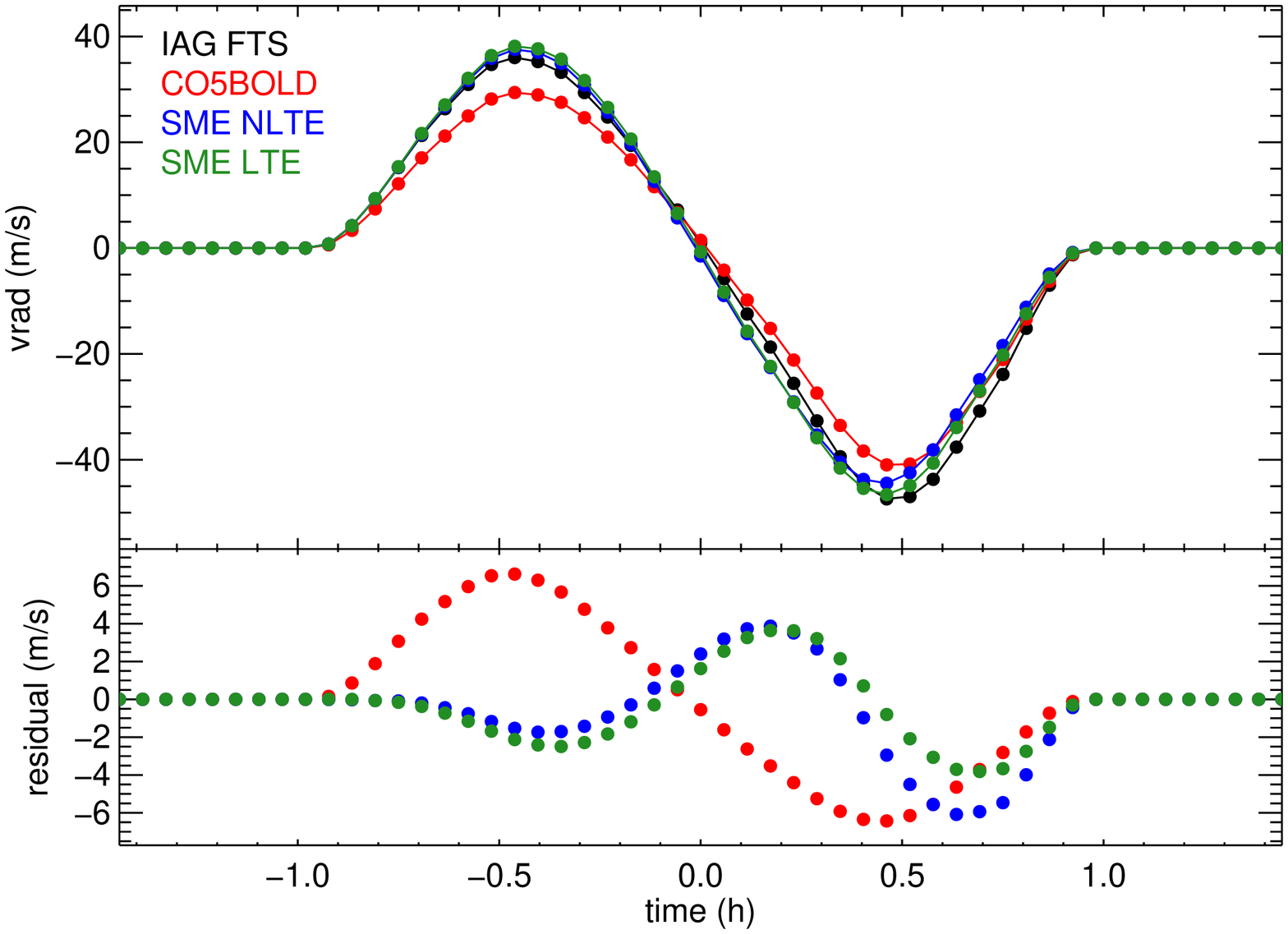}
  \mbox{\parbox{.95\textwidth}{\small \hspace{70mm} \ion{Ca}{i}}}\\[-5mm]
  \includegraphics[width=.9\hsize, viewport=8 5 625 450]{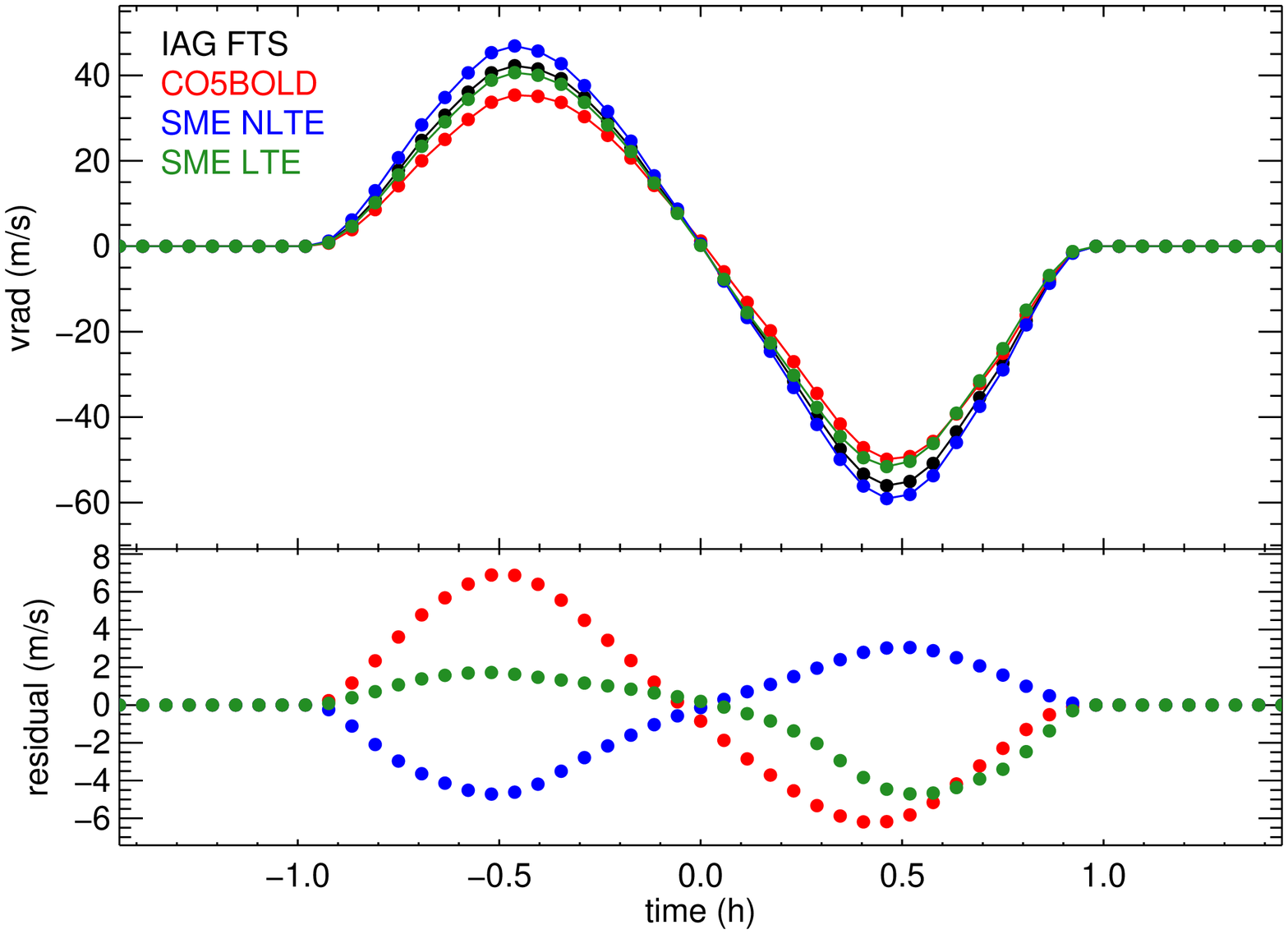}
  \mbox{\parbox{.95\textwidth}{\small \hspace{70mm} \ion{Fe}{i}}}\\[-5mm]
  \includegraphics[width=.9\hsize, viewport=8 5 625 450]{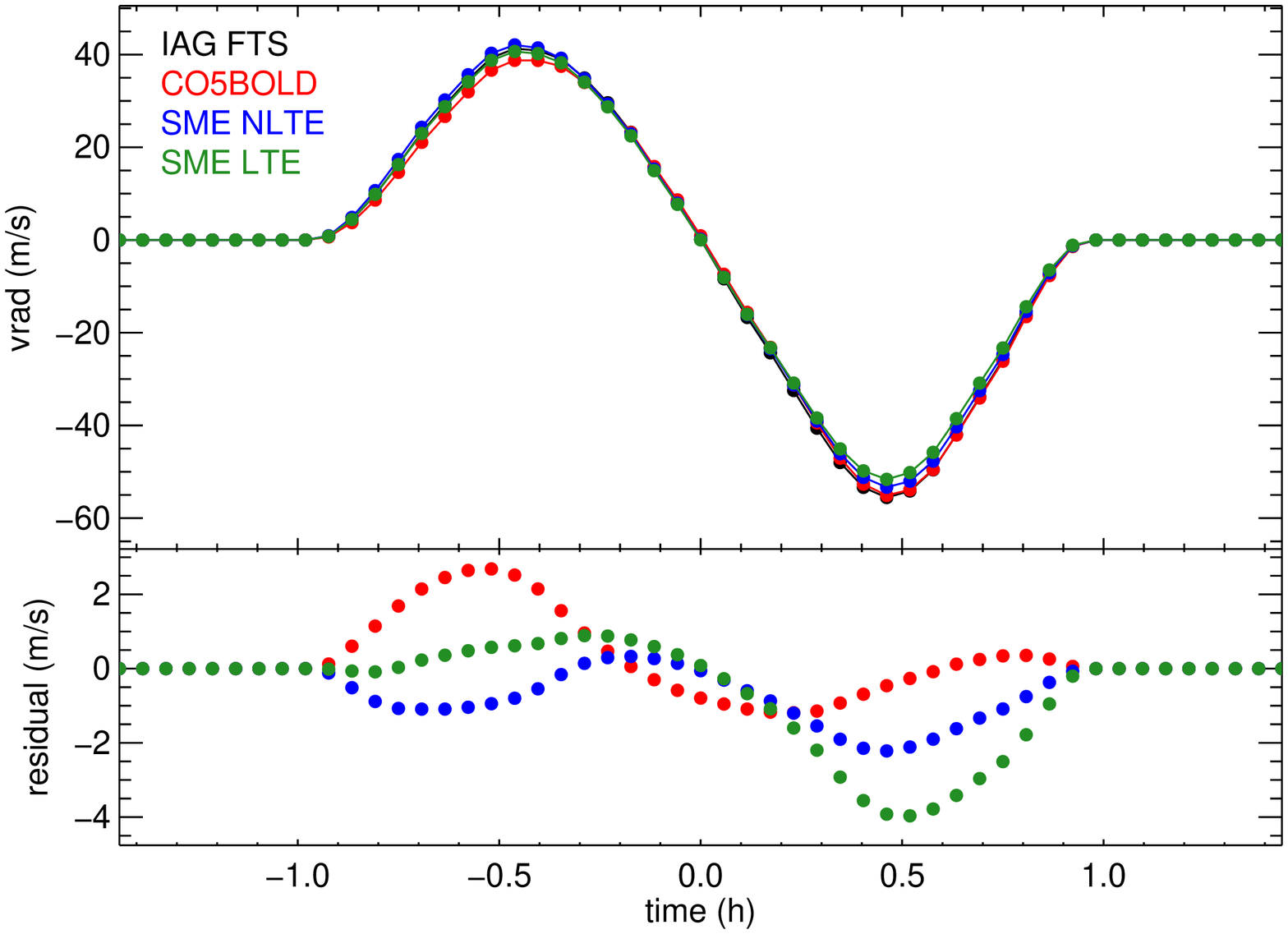}
  \caption{\label{fig:RM}Example RM curves and residuals between IAG FTS and
    model transmission curves for \ion{Na}{i}, \ion{Mg}{i}, \ion{Ca}{i}, and
    \ion{Fe}{i} (top to bottom). The configuration corresponds to the
    HD~189733b system (see Table\,\ref{tab:configuration}).}
  }
\end{figure}

A wavelength dependence of planetary radius causes different amplitudes of the
RM signal at different wavelengths, which is roughly proportional to the
star's $\varv \sin{i}$ and the square of the planet-to-star radius ratio
\citep{2005ApJ...622.1118O}. While this can be used to measure the planetary
transmission spectrum \citep{2004MNRAS.353L...1S, 2009A&A...499..615D}, a
chromatic RM effect can also be the result from chromatic CLV. The apparent
variability of the line profile centroid during transit, visible in the RM
curves, is shown for four spectral lines in Fig.\,\ref{fig:RM}. For each line,
we simulated the disk-integrated line profiles with our observations of the
solar surface and with the model line profiles from SME LTE, SME NLTE, and
CO$^5$BOLD. We computed the line center (the apparent radial velocity) by
fitting a Gauss function to the cross-correlation between the disk-integrated
spectra and the out-of-transit spectrum. Residuals between RM curves using
solar observations and each of the model profiles are shown in the lower panel
for each line.

The apparent radial velocities during transit depend on the shape of the line
profiles, and also on the definition of line center for asymmetric
lines. Here, we use the center of a Gauss fit to the cross-correlation between
in- and out-of-transit spectra. In our examples, the overall shape of the RM
curves are similar between the different spectral lines, but their amplitudes
vary between $\sim$60\,m\,s$^{-1}$ in the \ion{Na}{i} line and
$\sim$100\,m\,s$^{-1}$ in the \ion{Fe}{i} line. This is because the intrinsic
line profiles and their CLV are different and therefore the occultation of the
stellar surface at the same projected local velocity leads to different line
profile deformations in different spectral lines.

In general, the four versions of local line profiles lead to consistent RM
curves for each of the four spectral lines, albeit the \cobold profiles tend
to show an amplitude up to 10\,m\,s$^{-1}$ weaker than the other three sets of
profiles. The SME LTE and NLTE calculations lead to relatively similar
outcomes, especially for the \ion{Mg}{i} and the \ion{Fe}{i} lines. The
deviations between the RM curves computed with IAG FTS profiles and those from
the simulations show no preferred pattern of symmetry, except for the \cobold
case in which the RM curve is always underestimated before midtransit. This
is likely caused by the overpredicted blue wing of the \cobold calculations
(see Section\,\ref{sect:coboldcalc}). During the first half of the transit,
the planet occults the blueshifted areas of the rotating stellar surface,
which leads to a smaller RM amplitude if the blue wing is broader. We find
that for the HD\,189733 configuration, the uncertainty in the RM curve due to
unknown CLV is on the order of a few m\,s$^{-1}$ and always smaller than
10\,m\,s$^{-1}$.

\begin{figure}
  \centering
  \resizebox{\hsize}{!}{\includegraphics[viewport=15 5 640 450]{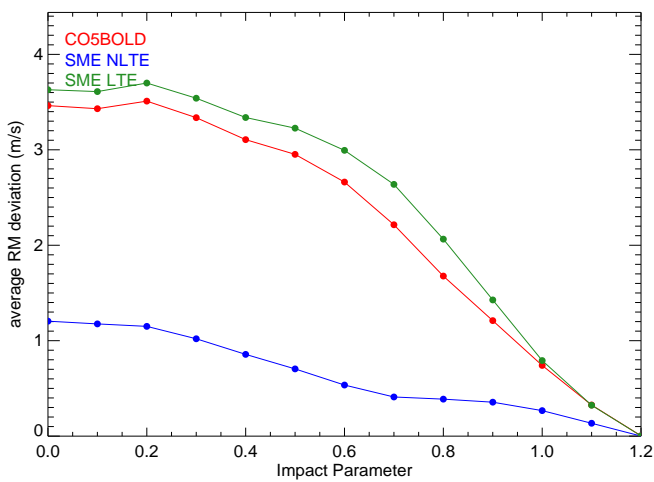}}
  \caption{\label{fig:ImpactStudyRM}Impact study showing average RM deviation
    change for the \ion{Na}{i} line depending on the impact parameter,
    $a$. The configuration is the same as in Table\,\ref{tab:configuration}, but with orbital
    inclination according to the value of $a$.}
\end{figure}

We investigate the mismatch between RM curves from the model line profiles and
the observed solar CLV for different transit configurations in an impact
parameter study. In Fig.\,\ref{fig:ImpactStudyRM}, we show the average
deviation (with the same definition as for the average transmission curve in
Section\,\ref{sect:transmissioncurve}) during transit between model and solar
CLV for the three models as a funcion of the impact parameter for the \ion{Na}{i}
line. As expected from the HD~189733b simulation shown in the top panel of
Fig.\,\ref{fig:RM}, the SME NLTE profiles show the smallest deviation from the
solar CLV case, but this would be different for other lines. The mismatch
tends to to be larger when the transit happens closer to the stellar center
and reaches average values exceeding 3\,m\,s$^{-1}$.

\section{Summary and discussion}

We present a detailed investigation of CLV observed in the solar library from
\citet{Ellwarth2022}. We focus on four spectral lines in the visible
wavelength range from four different ions: \ion{Na}{i}, \ion{Mg}{i},
\ion{Ca}{i}, and \ion{Fe}{i}. We compare the line shape and CLV to three different
spectral models. Our first synthetic spectrum was computed under the assumption
of LTE and without convection (SME LTE), the second included NLTE effects but
no convection (SME NLTE), and the third included 3D convection but line
formation is based on LTE (CO$^5$BOLD). Our goal was to separate the effects of
NLTE and convection on spectral line formation and CLV, and to estimate to
which extent the models can reproduce the behavior of spectral lines.

In most cases, the line depths are best matched by the NLTE model with the
exception of \ion{Mg}{i}. The LTE models typically produce lines that are
weaker than the observations, and the CLV is generally overestimated by the
models. All four example lines are relatively strong with core transmission
being less than 20\,\%. They show little CLV in the line cores but strong
differences with limb position in the wings of the lines where CLV is
relatively better predicted by the models. For the \ion{Na}{i} line,
neglecting NLTE effects produces a particularly large difference between the model
and observations.

The observed spectral lines are asymmetric, and their shape varies with limb
position. Lines from 1D models are symmetric by design and
therefore cannot reproduce line asymmetries and their limb effect. The \cobold
calculations appear more asymmetric than the observation, and they tend to
overpredict the blue wings of the line profiles. The relative variability of the
line shape with limb position predicted by the \cobold model shows trends
similar to the observed CLV, but its variability comes out stronger than
observed. Qualitatively, the best match is achieved for the \ion{Fe}{i} line,
in which NLTE effects are negligible and the \cobold model reproduces the
observations relatively well.

We show simulations of spectral lines during planetary transits and
investigate the impact of CLV on the planetary trace caused by occultation
only, that is, the RM effect, for a Sun-like star. This is relevant for
understanding the influence of CLV on the detectability of planetary
atmospheres. We find that narrow-band transmission curves are typically
overestimated by any of the three models; predictions of transit depth are up
to a factor of two larger than the ones computed with the observed line
profiles. In the RM curves, the models predict amplitudes that can be lower or
larger in amplitude than the curves based on solar observations. The maximum
deviations we found for a transit with a configuration similar to HD~189733b
are roughly 5\,m\,s$^{-1}$ for the \ion{Na}{i} line. Peak amplitude
differences are on the order of 5--10\,m\,s$^{-1},$ but they could be larger
for other lines and configurations, and in different stars. This can easily
exceed the chromatic RM expected from the presence of atoms such as Na in the
atmosphere of a planet \citep{2004MNRAS.353L...1S, 2015A&A...580A..84D,
  2020A&A...643A..64O}.

We conclude that neither of the three models provides a representative
prediction of any of the four strong lines used in our examples. This is not
surprising because the model spectra we used were not optimized to match the
observed line profiles. The solar observations provide important information
for atmospheric and convection models. Our simulations of transmission curves
can be used to estimate the uncertainty in analyses of exoplanet atmosphere
spectra for spectral lines that appear in the stellar spectrum. For Sun-like
stars, models seem to overpredict the signature of the planetary trace that
occurs in the absence of a planetary atmosphere. Extrapolations to stars
different than the Sun are not straightforward because line formation
processes can be very different in their atmospheres. We provide data and
models used here to allow simulations of planetary transits in the spectral
lines contained in the data.

\begin{acknowledgements}
  We thank T.~Olander for helpful information about runing SME in NLTE mode,
  and M.~Zechmeister for careful revision of the manuscript. We thank the
  referee for very helpful suggestions.
\end{acknowledgements}
  
\bibliographystyle{aa}
\bibliography{refs}

\appendix

\section{Abundances used in the synthetic spectrum calculations}
\label{app:abundances}

\begin{table}[h]
  \centering
  \caption{Chemical abundances used in the computation of the SME spectra and
    the synthetic 3D spectrum based on the \cobold\ model atmosphere.}
  \label{t:cobold-abund}
  \begin{tabular}{rlrr|rlrr}
    \hline\hline\noalign{\smallskip}
    Z  &   X  &   \multicolumn{2}{c}{$\log {\rm A(X})$}   &  Z  &   X  &   \multicolumn{2}{c}{$\log {\rm A(X})$}\\
       &      &   SME & \cobold &  &  & SME & \cobold \\
    \hline\noalign{\smallskip}
    1  &  H   &     12.00 & 12.00   &  44  &  Ru  &     1.84 &    1.84  \\
    2  &  He  &     10.93 & 10.93   &  45  &  Rh  &     1.12 &    1.12  \\
    3  &  Li  &      1.05 &  1.10   &  46  &  Pd  &     1.66 &    1.69  \\
    4  &  Be  &      1.38 &  1.40   &  47  &  Ag  &     0.94 &    0.94  \\
    5  &  B   &      2.70 &  2.55   &  48  &  Cd  &     1.77 &    1.77  \\
    6  &  C   &      8.39 &  8.52   &  49  &  In  &     1.60 &    1.66  \\
    7  &  N   &      7.78 &  7.92   &  50  &  Sn  &     2.00 &    2.00  \\
    8  &  O   &      8.66 &  8.83   &  51  &  Sb  &     1.00 &    1.00  \\
    9  &  F   &      4.56 &  4.56   &  52  &  Te  &     2.19 &    2.24  \\
    10 &  Ne  &      7.84 &  8.08   &  53  &  I   &     1.51 &    1.51  \\
    11 &  Na  &      6.17 &  6.33   &  54  &  Xe  &     2.24 &    2.17  \\
    12 &  Mg  &      7.53 &  7.58   &  55  &  Cs  &     1.07 &    1.13  \\
    13 &  Al  &      6.37 &  6.47   &  56  &  Ba  &     2.17 &    2.13  \\
    14 &  Si  &      7.51 &  7.55   &  57  &  La  &     1.13 &    1.17  \\
    15 &  P   &      5.36 &  5.45   &  58  &  Ce  &     1.70 &    1.58  \\
    16 &  S   &      7.14 &  7.33   &  59  &  Pr  &     0.58 &    0.71  \\
    17 &  Cl  &      5.50 &  5.50   &  60  &  Nd  &     1.45 &    1.50  \\
    18 &  Ar  &      6.18 &  6.40   &  62  &  Sm  &     1.00 &    1.01  \\
    19 &  K   &      5.08 &  5.12   &  63  &  Eu  &     0.52 &    0.51  \\
    20 &  Ca  &      6.31 &  6.36   &  64  &  Gd  &     1.11 &    1.12  \\
    21 &  Sc  &      3.17 &  3.17   &  65  &  Tb  &     0.28 &   $-$0.10  \\
    22 &  Ti  &      4.90 &  5.02   &  66  &  Dy  &     1.14 &    1.14  \\  
    23 &  V   &      4.00 &  4.00   &  67  &  Ho  &     0.51 &    0.26  \\
    24 &  Cr  &      5.64 &  5.67   &  68  &  Er  &     0.93 &    0.93  \\
    25 &  Mn  &      5.39 &  5.39   &  69  &  Tm  &     0.00 &    0.00  \\
    26 &  Fe  &      7.45 &  7.50   &  70  &  Yb  &     1.08 &    1.08  \\
    27 &  Co  &      4.92 &  4.92   &  71  &  Lu  &     0.06 &    0.06  \\
    28 &  Ni  &      6.23 &  6.25   &  72  &  Hf  &     0.88 &    0.88  \\
    29 &  Cu  &      4.21 &  4.21   &  73  &  Ta  &  $-$0.17 &   $-$0.13  \\
    30 &  Zn  &      4.60 &  4.60   &  74  &  W   &     1.11 &    1.11  \\
    31 &  Ga  &      2.88 &  2.88   &  75  &  Re  &     0.23 &    0.28  \\
    32 &  Ge  &      3.58 &  3.41   &  76  &  Os  &     1.25 &    1.45  \\
    33 &  As  &      2.29 &  2.37   &  77  &  Ir  &     1.38 &    1.35  \\
    34 &  Se  &      3.33 &  3.41   &  78  &  Pt  &     1.64 &    1.80  \\
    35 &  Br  &      2.56 &  2.63   &  79  &  Au  &     1.01 &    1.01  \\
    36 &  Kr  &      3.25 &  3.31   &  80  &  Hg  &     1.13 &    1.13  \\
    37 &  Rb  &      2.60 &  2.60   &  81  &  Tl  &     0.90 &    0.90  \\
    38 &  Sr  &      2.92 &  2.97   &  82  &  Pb  &     2.00 &    1.95  \\
    39 &  Y   &      2.21 &  2.24   &  83  &  Bi  &     0.65 &    0.71  \\
    40 &  Zr  &      2.58 &  2.60   &  90  &  Th  &     0.06 &    0.09  \\
    41 &  Nb  &      1.42 &  1.42   &  92  &  U   &  $-$0.52 &   $-$0.50  \\
    42 &  Mo  &      1.92 &  1.92   &      &      &          &  \\

    \hline\noalign{\smallskip}
  \end{tabular}
\end{table}

\end{document}